\documentclass[useAMS,usenatbib]{mn2e}
\usepackage{graphicx}    % Include figure files
\usepackage{dcolumn}     % Align table columns on decimal point
\usepackage{bm}          % bold math
\usepackage{amssymb,amsmath}  
\usepackage{color}
\usepackage{fixltx2e}
\usepackage{hyperref}

\newcommand{\nv}{\hat{\bf n}}

\title[Blind foreground subtraction for intensity mapping experiments]
      {Blind foreground subtraction for intensity mapping experiments}
\author[D. Alonso et al.]
  {David Alonso$^1$\thanks{E-mail:david.alonso@astro.ox.ac.uk}, Philip Bull$^{2}$, Pedro G. Ferreira$^1$,
   M\'ario G. Santos$^{3,4,5}$\\
   $^1$ Astrophysics, University of Oxford, DWB, Keble Road, Oxford OX1 3RH, UK\\
   $^2$ Institute of Theoretical Astrophysics, University of Oslo, P.O. Box 1029 Blindern, N-0315 Oslo, Norway\\
   $^3$ Department of Physics, University of Western Cape, Cape Town 7535, South Africa\\
   $^4$ SKA SA, 3rd Floor, The Park, Park Road, Pinelands, 7405, South Africa\\
   $^5$ CENTRA, Instituto Superior T\'ecnico, Universidade de Lisboa, Lisboa 1049-001, Portugal\\
  }
\begin{document}
  \date{\today}
  \pagerange{1--11} \pubyear{2014}
  \maketitle
  %\label{firstpage}

  \begin{abstract}
    We make use of a large set of fast simulations of an intensity mapping experiment
    with characteristics similar to those expected of the Square Kilometre Array (SKA)
    in order to study the viability and limits of blind foreground subtraction
    techniques. In particular, we consider three different approaches: polynomial
    fitting, principal component analysis (PCA) and independent component analysis
    (ICA). We review the motivations and algorithms for the three methods, and
    show that they can all be described, using the same mathematical framework, as
    different approaches to the blind source separation problem. We study the
    efficiency of foreground subtraction both in the angular and radial (frequency)
    directions, as well as the dependence of this efficiency on different instrumental
    and modelling parameters. For well-behaved foregrounds and instrumental effects we
    find that foreground subtraction can be successful to a reasonable level on most
    scales of interest. We also quantify the effect that the cleaning has on the recovered
    signal and power spectra. Interestingly, we find that the three methods yield
    quantitatively similar results, with PCA and ICA being almost equivalent.
  \end{abstract}

%--------------------------------------------------------------------------------------

  \section{Introduction}\label{sec:intro}
    The last few decades have seen a radical improvement in the quantity and quality
    of observational data that has transformed cosmology into a fully-fleshed data-driven 
    science --
    so much so that it has now become common to refer to the current status of the field as
    the ``era of precision cosmology''. The vast majority of these data come from two
    very distinct regimes, both observationally and physically: on the one hand, observations
    of the cosmic microwave background (CMB), emitted in the early Universe
    ($z\sim1000$), have allowed us to measure the values of most cosmological parameters with
    astonishing precision \citep{2013ApJS..208...19H,2013arXiv1303.5076P}. On the other,
    astronomical observations in the optical range of wavelengths have supplied a wealth of data
    with which we have been able to characterise the late-time ($z\lesssim1$) evolution of the
    Universe and to make one of the most puzzling discoveries in cosmology: its accelerated
    expansion \citep{Riess:1998cb,Perlmutter:1998np}. In this regime, galaxy redshift surveys
    \citep{2014MNRAS.441...24A,2013MNRAS.430..924C} have allowed us to draw a plausible picture
    describing the evolution of the primordial density perturbations observed in the CMB to form
    the non-linear large-scale structure (LSS) that we observe today.
    
    There exist, however, multiple open questions in cosmology, the study of which would 
    greatly benefit from observational data in the remaining intermediate range of redshifts.
    For instance, one of the most important stages in the evolution of the Universe, the
    epoch of reionisation (EoR) and the formation of the first stars and galaxies, must
    occur at $z\sim10$, but its precise nature remains unknown due to the lack of observations
    \citep{2006MNRAS.365..115F}. Furthermore, being able to observe the distribution of 
    matter out to
    redshift $z\simeq3-4$ would make a very large volume available for LSS analyses, which could
    significantly improve our understanding of the process of structure formation, as well as
    opening the field to the study of ultra-large scales, which contain substantial 
    information about
    the underlying gravitational theory and the statistics of the primordial density field
    \citep{2013PhRvD..87f4026H,2013PhRvL.111q1302C}. Radio-astronomical observations offer an
    excellent tool in this regard through the measurement of the 21cm signal from neutral hydrogen
    (HI). Measuring the intensity of this line emission, due to the hyperfine spin-flip transition
    ($\nu_{21}\simeq1420\,{\rm MHz}$), as a function of frequency constitutes an ideal way to
    trace the HI density field as a function of redshift $z=\nu_{21}/\nu-1$.
    
    While the potential scientific value of these techniques is enormous, the detection of
    the HI signal faces several observational challenges. After reionisation, most of
    the neutral hydrogen in the Universe is thought to reside in high density regions inside
    galaxies, in what are known as damped Lyman-$\alpha$ systems, where it is shielded from
    the ionising UV photons. Unfortunately, the HI emission from individual galaxies is
    usually too weak to be measured efficiently, and thus a technique known as {\sl intensity
    mapping} has been advocated \citep{Battye:2004re,2006ApJ...653..815M,Chang:2007xk,
    2008PhRvD..78b3529M,2008PhRvD..78j3511P,2008MNRAS.383..606W,2008MNRAS.383.1195W,
    2008PhRvL.100p1301L,2009astro2010S.234P,Bagla:2009jy,Seo:2009fq,2011ApJ...741...70L,
    2012A&A...540A.129A,2013MNRAS.434.1239B,2005ApJ...624L..65B,2013ApJ...763L..20M,
    2013MNRAS.434L..46S,2014arXiv1405.1452B}, in which the combined emission from
    all sources in wide angular pixels is observed instead. While small angular
    scales are lost through this method, it is possible to retain the larger scales of most
    cosmological relevance, such as the baryon acoustic oscillation (BAO) scale. Thus
    intensity mapping constitutes a promising method to efficiently observe the
    distribution of matter in the Universe on cosmological scales.
    
    The success of intensity mapping relies on being able to separate the cosmological
    HI signal from that of foreground radio sources emitting in the same frequency range, however.
    The most important of these sources, synchrotron emission from our own galaxy, is
    $\sim5$ orders of magnitude larger than the expected 21cm signal, even at high galactic
    latitudes, and so the problem of foreground subtraction is of prime importance.
    This topic has been widely covered in the literature \citep{2002ApJ...564..576D,
    2003MNRAS.346..871O,2005ApJ...625..575S,2006ApJ...648..767M,2006ApJ...650..529W,
    2008MNRAS.391..383G,2008MNRAS.389.1319J,2009A&A...500..965B,2010A&A...522A..67B,
    2010MNRAS.409.1647J,2013ApJ...769..154M}, mostly for the EoR regime, and several
    foreground removal techniques have been proposed \citep{2009MNRAS.398..401L,
    2011PhRvD..83j3006L,2013ApJ...763L..20M,2014MNRAS.441.3271W,2014ApJ...781...57S,
    2014arXiv1401.2095S}. Fortunately, as opposed to the cosmological signal, most relevant
    foregrounds have a very smooth frequency dependence, a fact that can be exploited to
    subtract them efficiently. Nevertheless, their large amplitude makes a thorough study
    of the consequences of foreground cleaning on the recovered cosmological signal indispensable.
    
    Foreground removal techniques can be classified as being either {\sl blind}, if they
    only make use of generic foreground properties like spectral smoothness, or
    {\sl non-blind}, if a more detailed model of the foregrounds is required. In this
    paper we have used a large set of simulations of both the cosmological signal and
    several different foregrounds to study and compare the performance of different blind
    foreground subtraction methods. The paper is structured as follows: in Section
    \ref{sec:imap} we briefly describe the physics of the HI cosmological signal and the most
    relevant foregrounds for intensity mapping. Section \ref{sec:bfg} covers the mathematics of
    blind foreground cleaning within a unified formalism. Section \ref{sec:method} contains the
    details of the simulations used in this analysis, as well as the criteria used to compare the
    different methods. Finally, Section \ref{sec:results} presents the results of this study
    regarding the performance and limits of blind foreground cleaning, and Section
    \ref{sec:discussion} summarises our findings.
    
  \section{Intensity mapping and its foregrounds}\label{sec:imap}
    The emissivity (luminosity per unit frequency and solid angle) of a cloud of neutral
    hydrogen due to the 21cm line is given, in its rest frame, by
    \begin{equation}
      \frac{dL}{d\Omega\,d\nu}=\frac{h_p\,A_{21}}{4\pi}\frac{N_2}{N_{\rm H}}N_{\rm H}\,
      \nu\,\varphi(\nu),
    \end{equation}
    where $h_p$ is Planck's constant, $A_{21}=2.876\times10^{-15}\,{\rm Hz}$ is the Einstein
    coefficient corresponding to the 21cm line, $N_2$ is the number of atoms in the 
    excited state, $N_{\rm H}$ is the total number of HI atoms and $\varphi(\nu)$ is the
    normalised line profile. From this result it is easy to calculate the intensity 
    measured by an observer in the lightcone and hence the brightness temperature along a
    particular direction $\nv$ in the Rayleigh-Jeans approximation \citep{2005MNRAS.360...27A}:
    \begin{align}\nonumber
      T(z,\nv)&=\frac{3\,h_pc^3A_{21}}{32\pi k_B\nu_{21}^2}\frac{(1+z)^2}{H(z)}
      n_{\rm HI}(z,\nv)\\\label{eq:thiofz}
              &=(0.19055\,{\rm mK})\frac{\Omega_{\rm b}\,h\,(1+z)^2x_{\rm HI}(z)}
                {\sqrt{\Omega_{\rm M}(1+z)^3+\Omega_{\Lambda}}}(1+\delta_{\rm HI}),
    \end{align}
    where $n_{\rm HI}\propto(1+\delta_{\rm HI})$ is the comoving number density of HI atoms,
    $k_B$ is Boltzmann's constant and $x_{\rm HI}$ is the fraction of the total baryonic mass in
    HI. Note that to reach this result we have neglected scattering and self-absorption of the
    emitted photons, as well as any relativistic effects other than the Hubble expansion. 
    The observed HI signal is also affected by redshift-space distortions 
    \citep{Kaiser:1987qv} in the same way as galaxy number counts are \citep{2013PhRvD..87f4026H}.

    The most relevant foregrounds for intensity mapping can be classified as being extragalactic
    (caused by astrophysical sources beyond the Milky Way) or galactic. In terms of amplitude,
    the largest foreground by far is galactic synchrotron emission (GSE), caused by high-energy
    cosmic-ray electrons accelerated by the Galactic magnetic field. Although at the lowest
    frequencies the GSE spectrum is damped by self-absorption and free-free absorption, for the
    relevant radio frequencies it should be possible to describe it by a single power law
    $T_{\rm GSE}\propto\nu^\beta$, with a spectral index $\beta(\nv)\approx-2.8$
    \citep{2009ApJ...705.1607D,2013A&A...553A..96D}. GSE is expected to be partially linearly
    polarised, and so its polarised part will undergo Faraday rotation as it travels through the
    Galactic magnetic field and the interstellar medium \citep{1986rpa..book.....R,
    2009A&A...495..697W}. This effect gives polarised GSE a non-trivial and non-smooth
    frequency dependence, which would make it a very troublesome foreground if leaked into the
    unpolarised signal \citep{2010MNRAS.409.1647J,2013ApJ...769..154M,2014PhRvD..89l3002D}.
    
    Another source of foreground emission within our Galaxy is the so-called free-free emission,
    caused by free electrons accelerated by ions, which traces the warm ionised medium. As in the
    case of GSE, free-free emission is predicted to be spectrally smooth in the relevant range of
    frequencies \citep{2003MNRAS.341..369D}. All other known potential galactic foregrounds, such
    as anomalous microwave emission or dust emission, should be negligible below
    $\sim1\,{\rm GHz}$.
    
    Extragalactic radio sources can be classified into two different categories: bright radio
    galaxies, such as active galactic nuclei, and ``normal'' star-forming galaxies. While the
    distribution of the former should be Poisson-like, the clustering of the latter should
    be significant, which has a direct impact on the degree of smoothness of their joint
    emission \citep{2005ApJ...625..575S,2004ApJ...608..622Z}. Note that radio sources have been
    observed up to very high redshifts ($z\sim5$), and therefore many of them will be physically
    behind the cosmological signal in the case of intensity mapping. However we will refer to all
    sources of radio emission that need to be separated from the cosmological signal as
    ``foregrounds'', trusting that the use of this term will not cause any confusion.

    Other possible foreground sources are atmospheric noise, artificial radio frequency
    interference (which we discuss in section \ref{ssec:corrlength}) and line foregrounds, caused
    by line emission from astrophysical sources in other frequencies. Due to the spectral isolation
    of the 21cm line, together with the expected low intensity of the most potentially harmful
    lines (such as OH at $\nu_{\rm OH}\sim1600\,{\rm MHz}$), the HI signal should be very robust
    against line confusion.
    
  \section{Blind foreground subtraction}\label{sec:bfg}
    As has been noted above, the most relevant foregrounds for intensity mapping
    should be correlated (i.e. smooth) in frequency. Blind foreground subtraction
    methods make use of this property, modelling the sky brightness temperature
    along a given direction in the sky $\nv$ and at a frequency $\nu$ as
    \begin{equation}
      T(\nu,\nv)=\sum_{k=1}^{N_{\rm fg}}f_k(\nu)\,S_k(\nv)+
      T_{\rm cosmo}(\nu,\nv)+T_{\rm noise}(\nu,\nv),
    \end{equation}
    Where $N_{\rm fg}$ is the number of foreground degrees of freedom to subtract,
    $f_k(\nu)$ are a set of smooth functions of the frequency, $S_k(\nv)$ are 
    the foreground sky maps and $T_{\rm cosmo}$ and $T_{\rm noise}$ are the
    cosmological signal and instrumental noise components.
    
    For a particular line of sight $\nv$ measured in a discrete set of
    $N_{\nu}$ frequencies, this model can be written as a linear system:
    \begin{equation}\label{eq:bfg_eq}
      {\bf x}=\hat{\bf A}\cdot{\bf s}+{\bf r},
    \end{equation}
    where $x_i=T(\nu_i,\nv)$, $A_{ik}=f_k(\nu_i)$, $s_k=S_k(\nv)$ and
    we have grouped the cosmological signal and thermal noise in a single
    component $r_i=T_{\rm cosmo}(\nu_i,\nv)+T_{\rm noise}(\nu_i,\nv)$. The 
    problem of foreground subtraction then boils down to devising a method to
    determine $\hat{\bf A}$ and ${\bf s}$ so that ${\bf r}={\bf x}-
    \hat{\bf A}{\bf s}$ is recovered as accurately as possible.
    The three methods described below correspond to three different approaches
    to this problem.
    
    \subsection{Polynomial fitting}\label{ssec:polog}
      Probably the most intuitive (and na\"ive) way of approaching blind
      foreground subtraction is to choose an ad-hoc basis of smooth functions $f_k$
      that we think can describe the foregrounds and then find the foreground maps
      $s_k$ by least-squares fitting the model in Eq. \ref{eq:bfg_eq} to each
      line of sight. This is done by minimising the $\chi^2$,
      \begin{equation}\label{eq:chi2}
        \chi^2=({\bf x}-\hat{\bf A}\cdot{\bf s})^T\,\hat{\bf N}^{-1},
        ({\bf x}-\hat{\bf A}\cdot{\bf s})
      \end{equation}
      where $\hat{\bf N}$ is the covariance matrix of ${\bf r}$. The solution is
      \begin{equation}\label{eq:polog_sol}
        {\bf s}=(\hat{\bf A}^T\hat{\bf N}^{-1}\hat{\bf A})^{-1}
        \hat{\bf A}^T\hat{\bf N}^{-1}{\bf x}.
      \end{equation}
      In this analysis we have neglected the correlation between different frequency
      bands and therefore we have assumed a diagonal covariance $N_{ij}=\sigma_i^2\,
      \delta_{ij}$\footnote{This
      assumption is not exactly correct, since even for uncorrelated instrumental
      noise, the cosmological signal (which we have included as a noise-like
      contribution in this formalism) is correlated in real space, which should
      ideally be taken into account.}.
      
      It is important to note that the deviations from the smooth foregrounds
      are due not only to thermal noise, but also to the cosmological clustering
      signal. For this reason, in order to optimally fit for the foregrounds we
      have used the joint variance of both components:
      \begin{equation}\label{eq:var_tot}
        \sigma_i=\sqrt{\sigma_{{\rm noise},i}^2+\sigma_{{\rm cosmo},i}^2}.
      \end{equation}
      Evidently it is not possible to compute $\sigma_{\rm cosmo}$ from the
      cosmological signal before subtracting the foregrounds. In our analysis we
      have done this by using the foreground-free maps, which wouldn't be available
      in a real experiment; we can propose three alternative methods, however:
      \begin{itemize}
        \item Assuming a particular cosmological model it would be possible to quantify the
              clustering variance, at least at the linear level, as:
              \begin{equation}
                \sigma_{{\rm cosmo},i}^2=
                \int \frac{dk^3}{(2\pi)^3} |W_i({\bf k}_\perp,k_\parallel)|^2 P(k),
              \end{equation}
              where $P(k)$ is the power spectrum for the assumed model and $W_{{\rm pix},i}$
              is the Fourier-space window function describing the beam and pixel size and the 
              radial width for the $i-{\rm th}$ frequency bin.
        \item Again assuming a particular cosmological model, another possibility would be
              to simulate the cosmological signal and include instrumental effects to compute
              $\sigma_i^2$ from the simulations.
        \item Through a multi-step analysis: first the foregrounds are subtracted using only
              the variance due to the instrumental noise (or even no variance weighting at
              all). The joint variance $\sigma_i$ is then estimated from the cleaned maps
              and the foreground subtraction is repeated using this better estimate. The
              process can then be repeated until it converges. This method would be
              computationally more demanding, but it has the advantage of being
              model-independent.
      \end{itemize}

      Previous studies of foreground subtraction in intensity mapping (particularly
      for the EoR regime) have made use of this technique in log-log space, i.e.
      using the model:
      \begin{equation}
        \log(T(\nu,\nv))=\sum_{k=1}^{N_{\rm fg}}\alpha_k(\nv)\,f_k(\log(\nu)).
      \end{equation}
      Following what has been done by other groups 
      \citep{2006ApJ...650..529W,2012A&A...540A.129A},
      we used polynomials as basis functions
      in this study, i.e. $f_k(\log(\nu))=[\log(\nu)]^{k-1}$. Note that although very
      similar in spirit, this procedure does not adhere exactly to the model in Eq.
      \ref{eq:bfg_eq}, since the fitting is done in log-log space. This also implies
      that the fit weights $\sigma_i^{-2}$ must be translated from linear space:
      \begin{equation}
        \sigma_{\rm log}(T)\simeq \frac{\sigma_{\rm lin}}{T}.
      \end{equation}

    \subsection{Principal component analysis}\label{ssec:pca}
      Principal component analysis (PCA) applied to foreground subtraction seeks to
      make use of the main properties of the foregrounds (namely their large
      amplitude and smooth frequency coherence) to find both the foreground
      components $s_k$ \emph{and} an optimal set of basis functions $A_{ik}$
      at the same time. Here we present the motivation for PCA and its connection
      to the main equation for blind foreground subtraction (Eq. \ref{eq:bfg_eq}).
      
      It is easy to prove that the frequency-frequency covariance matrix of any
      component that is very correlated in frequency will have a very particular
      eigensystem: most of the information will be concentrated in a small set
      of very large eigenvalues, the other ones being negligibly small. As an extreme
      case, consider a completely correlated component, whose covariance matrix
      is $C_{ij}=1\,\,(\forall\, i,j\in[1,N_{\nu}])$, and whose eigenvalues are
      $\lambda_1=N_{\nu},\,\,\lambda_i=0\,i>1$. Thus we can attempt to subtract
      the foregrounds by eliminating from the measured temperature maps the
      components corresponding to the eigenvectors of the frequency covariance
      matrix with the $N_{\rm fg}$ largest associated eigenvalues. The explicit
      method is as follows:
      \begin{enumerate}
        \item Compute the frequency covariance matrix from the data by
              averaging over the available $N_{\theta}$ lines of sight:
              \begin{equation}
                C_{ij}=\frac{1}{N_{\theta}}\sum_{n=1}^{N_{\theta}}
                T(\nu_i,\nv_n)T(\nu_j,\nv_n).
              \end{equation}
        \item Diagonalise the covariance matrix:
	      \begin{equation}
                \hat{\bf U}^{T}\,\hat{\bf C}\,\hat{\bf U}=\hat{\bf \Lambda}\equiv
                {\rm diag}(\lambda_1,...,\lambda_{N_{\nu}}),
	      \end{equation}
	      Where $\lambda_i>\lambda_{i+1}\,\forall i$ are the eigenvalues of
	      $\hat{\bf C}$, and $\hat{\bf U}$ is an orthogonal matrix whose
	      columns are the corresponding eigenvectors.
	\item At this stage we identify $N_{\rm fg}$ eigenvalues corresponding to the
	      foregrounds as those that are much larger than the rest. Depending
	      on the frequency structure of the foregrounds and the different
	      instrumental effects this number will be more or less evident (see the
	      discussion in section \ref{ssec:nfg}). We then build the matrix
	      $\hat{\bf U}_{\rm fg}$ from the columns of $\hat{\bf U}$ corresponding
	      to these eigenvalues and model the brightness temperature for each
	      line of sight as
	      \begin{equation}\label{eq:pca_sys}
	        {\bf x}=\hat{\bf U}_{\rm fg}\,{\bf s}+{\bf r},
	      \end{equation}
	      which is analogous to Equation \ref{eq:bfg_eq}. The foreground maps
	      ${\bf s}$ are then found by projecting ${\bf x}$ on the basis of
	      eigenvectors of $\hat{\bf C}$, yielding:
	      \begin{equation}\label{eq:pca_sol}
	        {\bf s}=\hat{\bf U}_{\rm fg}^T\,{\bf x}.
	      \end{equation}
      \end{enumerate}
      Note that, since $\hat{\bf U}$ is orthogonal,
      $\hat{\bf U}_{\rm fg}^T\hat{\bf U}_{\rm fg}={\bf 1}$, and, except for the
      presence of the covariance $\hat{\bf N}$, Equations \ref{eq:pca_sys} and
      \ref{eq:pca_sol} are equivalent to Equations \ref{eq:bfg_eq} and
      \ref{eq:polog_sol}.
      
      In the presence of $\hat{\bf N}$, the frequency covariance must be computed
      using an inverse-variance scheme:
      \begin{equation}\label{eq:pca_invar}
        C_{ij}=\frac{1}{N_{\theta}}\sum_{n=1}^{N_{\theta}}
        \frac{T(\nu_i,\nv_n)}{\sigma_i}\frac{T(\nu_j,\nv_n)}{\sigma_j},
      \end{equation}
      which is equivalent to carrying out the steps above using the weighted maps
      $x_i=T(\nu_i)/\sigma_i$ and multiplying by $\sigma_i$ in the end to obtain
      the de-weighted temperature maps. It is easy to see that doing this we introduce
      the missing $\hat{\bf N}^{-1}$ factors that make the equivalence between 
      Equations \ref{eq:pca_sol} and \ref{eq:polog_sol} complete.
      
      We thus see that the PCA method is in fact equivalent to the polynomial
      fitting method, with the set of basis functions $A_{ik}$ given by the data
      themselves in the form of those that that contain most of the total variance
      (i.e. the principal eigenvectors of the covariance matrix).
      
      Finally, let us note that in a real experiment the situation will
      be more complicated, for example due to the presence of correlated instrumental
      noise. This problem was addressed in the pioneering analysis done by the 
      Green Bank Telescope team \citep{2013ApJ...763L..20M,2013MNRAS.434L..46S} by
      computing the frequency covariance matrix through
      a cross correlation of temperature maps corresponding to different seasons.
      However this yields a non-symmetric covariance matrix, and therefore its
      singular value decomposition (SVD) must be used, instead of the PCA. The
      simulations used for this analysis, however, do not contain correlated noise,
      and hence we will not worry about these complications. We leave the
      analysis of such instrumental issues for future work.

    \subsection{Independent component analysis}\label{ssec:ica}
      Independent component analysis (ICA) tries to solve the blind equation
      (Eq. \ref{eq:bfg_eq}) under the assumption that the sources ${\bf s}$ are statistically
      independent from each other ($P({\bf s})=\prod_i P_i(s_i)$). Explicitly this is
      enforced by using the central limit theorem, according to which, if ${\bf x}$ is
      made up of a linear combination of linearly independent sources, its distribution
      should be ``more Gaussian'' than those of the independent sources. Therefore we
      can attempt to impose statistical independence by maximising any statistical
      quantity that describes non-Gaussianity.
      
      {\tt FastICA} \citep{fica1} is a relatively popular and computationally efficient
      algorithm to apply ICA to a general system. Here we will outline the operations
      carried out by {\tt FastICA} as well as its similarities with PCA. As before, we label
      the brightness temperature measured in different frequencies along a given line
      of sight $\nv$ by a vector ${\bf x}$, however the reader must bear in mind
      that {\tt FastICA} is provided with a number of samples of ${\bf x}$ (i.e. different
      pixels), which it uses to compute the expectation values in the equations below.

      {\tt FastICA} begins by ``whitening'' the data ${\bf x}$. This implies first of
      all performing a full PCA analysis of the data, decorrelating them and sorting
      them by the magnitude of the covariance eigenvalues. The uncorrelated variables
      are then divided by their standard deviations so as to impose a unit variance
      on all of them (this is done in order to simplify the subsequent steps in the
      algorithm). We are then left with the corresponding equation for the whitened data
      $\tilde{\bf x}\equiv\hat{\bf \Lambda}^{-1/2}\,\hat{\bf U}\,{\bf x}$:
      \begin{equation}\label{eq:fica}
        \tilde{\bf x}=\hat{\bf A}'\,{\bf s}\equiv
        \hat{\bf \Lambda}^{-1/2}\,\hat{\bf U}\,\hat{\bf A}\,{\bf s},
      \end{equation}
      where as before $\hat{\bf U}$ and $\hat{\bf \Lambda}$ are the eigenvectors
      and eigenvalues of the data covariance matrix. Note that although the data have
      been whitened, {\tt FastICA} still preserves their PCA order.
      The problem is further simplified by requiring that the sources be also ``white''
      (i.e., uncorrelated and unit variance), since that implies that $\hat{\bf A}'$ must
      be orthonormal.
      
      {\tt FastICA} then tries to find the independent components by inverting Equation
      \ref{eq:fica}:
      \begin{equation}
        {\bf s}=\hat{\bf W}\,\tilde{\bf x}.
      \end{equation}
      and finding the rows of $\hat{\bf W}$ by maximising the non-Gaussianity of the
      individual components. {\tt FastICA} parametrises the level of non-Gaussianity
      in terms of the negentropy $J(y)=H(y_{\rm G})-H(y)$, where $H(y)$ is the
      entropy of the variable $y$ and $y_{\rm G}$ is a unit-variance Gaussian random
      variable. Since for all variables with the same variance the entropy is maximal
      for a Gaussian variable, the negentropy $J$ is always positive for non-Gaussian
      variables. However, computing the negentropy requires an intimate knowledge of
      the probability distribution, which in general we lack. For this reason {\tt FastICA}
      uses an approximation to $J$ given by
      \begin{equation}
        J(y)\sim\sum_{i} k_i\left[\langle G_i(y)\rangle_{\theta}-
                                  \langle G_i(y_{\rm G})\rangle_{\theta}\right],
      \end{equation}
      where ($k_i$) are positive constants, $\langle\cdot\rangle_{\theta}$ denotes averaging over
      all available samples (i.e. pixels) and $G_i$ is a set of non-quadratic functions.
      {\tt FastICA} makes use of two functions in particular\footnote{No significant
      difference in the final foreground-cleaned maps was found for either choice of
      function.}:
      \begin{equation}
        G(y)=e^{-y^2/2},\hspace{12pt}G(y)=\frac{1}{a}\log\cosh(a\,y),\,\,1\leq a\leq 2.
      \end{equation}

      The rows of $\hat{\bf W}$ are found as follows:
      \begin{enumerate}
       \item First an initial unit random vector ${\bf w}$ is chosen as one of the rows
             of $\hat{\bf W}$ and its projection on the data $y\equiv {\bf w}^T{\bf x}$
             is computed.
       \item The correct direction of ${\bf w}$ is found by maximising the negentropy
             of $y$. This is done by an iterative Newton-Raphson algorithm (renormalising
             the modulus of ${\bf w}$ after each step).
       \item The process ends when a given Newton-Raphson iterations yields a vector that
             is close enough to the one from the previous step.
      \end{enumerate}
      Further details regarding the algorithm and its relation to other existing ICA
      methods can be found in \citet{fica1,fica2}
      
      It is important to note that the second step in {\tt FastICA}'s algorithm (namely,
      the maximisation of the negentropy) will be unable to separate Gaussian foreground
      components beyond the initial PCA, since their negentropy is identically 0.
      This is simply a consequence of the fact that ICA is equivalent to PCA for
      the case in which all the sources are Gaussian\footnote{It is a common misconception
      to explain the ability of ICA to remove radio foregrounds by saying that it searches for
      non-Gaussianity in the frequency direction, and thus separates the ``noisy'' components
      (cosmological signal and instrumental noise) from the smooth foregrounds. It is
      important to stress the fact that non-Gaussianity is in reality maximised
      in angles and not in frequency (i.e. the negentropy is estimated by averaging over
      pixels), and that the motivation for this is to improve on the initial PCA separation
      by imposing statistical independence, rather than to distinguish noisy and smooth
      components.}.
      
      Unfortunately, as described in section \ref{ssec:sims}, the foreground maps corresponding
      to point sources and free-free emission were generated as Gaussian random fields in
      our simulations, and therefore we will not be able to appreciate the full power of
      ICA in removing those. However, the bulk of the galactic synchrotron emission
      (which is the most relevant foreground for intensity mapping) was extrapolated from
      the Haslam map \citep{1982A&AS...47....1H}, and is therefore non-Gaussian. Thus we
      should still be able to evaluate the impact of ICA methods in foreground removal.
      
    \subsection{Summary}\label{ssec:fg_sum}
      As we have shown in the previous three sections, the three blind methods under
      study in this paper can be understood as three different ways of approaching the
      least-squares problem of Equations \ref{eq:bfg_eq} and \ref{eq:polog_sol}, each following
      different criteria in order to find the basis functions $\hat{\bf A}$.
      \begin{itemize}
       \item {\bf Line-of-sight (polynomial) fitting} imposes an ad-hoc set of basis functions
             that, we think, should be able to describe the foregrounds accurately enough.
       \item {\bf PCA} finds $\hat{\bf A}$ by requiring that the sources should be
             uncorrelated and account for the majority of the total variance in the data.
       \item {\bf ICA}, like PCA, maximises the total variance but instead of
             \emph{uncorrelatedness} it imposes \emph{statistical independence}. Since
             these two properties are equivalent for Gaussian variables, PCA and ICA are
             mathematically equivalent when all the sources are Gaussian.
      \end{itemize}

  \section{Method}\label{sec:method}
    In order to compare the three blind methods discussed in the previous section we have
    tested them on a suite of 100 fast simulations of the cosmological signal and most
    relevant foregrounds. In this section we will describe the characteristics of these
    simulations as well as the criteria used to compare the different methods.
    
    \subsection{The simulations}\label{ssec:sims}
      The simulations used here were generated using the publicly available code presented
      in \citet{2014arXiv1405.1751A}
      \footnote{\url{http://intensitymapping.physics.ox.ac.uk/CRIME.html}}. The code generates
      random realisations of the cosmological signal and most relevant foregrounds, and 
      includes a basic set of instrumental effects.
      
      \paragraph*{The cosmological signal.} We generate the cosmological signal by producing a
        three-dimensional Gaussian realisation of the matter density and velocity fields in a
        cubic grid. The Gaussian density field is transformed into a more physically motivated
        one using the log-normal transformation \citep{Coles:1991if}. During this step we also
        apply a linear redshift-dependent bias and lightcone evolution based on linear
        perturbation theory to the overdensity field. This field is interpolated onto a set of
        temperature maps at different frequencies (i.e. redshifts) using Equation \ref{eq:thiofz}
        At this point, redshift-space distortions are included using the radial velocity field.
        For these simulations we used the parametrisation $x_{\rm HI}=0.008\,(1+z)$ for the
        neutral hydrogen fraction, based on the observations of \citet{2005MNRAS.359L..30Z} and
        \citet{2005ARA&A..43..861W}, and for simplicity we assumed that HI is an unbiased tracer
        of the density field ($\delta_{\rm HI}=\delta$). Further details regarding this method
        can be found in \citet{2014arXiv1405.1751A}.
        
        A box of size $L_{\rm box}=8240\,{\rm Mpc}/h$ was used with a Cartesian grid of
        2048 cells per side. This way we are able to generate full-sky maps to redshift
        $z_{\rm max}=2.55$ with a spatial resolution of $\Delta x\simeq 4\,{\rm Mpc}/h$.
        The cosmological signal was generated using Planck-compatible parameters:
        $(\Omega_M,\Omega_b,\Omega_k,h,w_0,w_a,\sigma_8,n_s)=(0.3,0.049,0,0.67,-1,0,0.8,0.96)$,
        and the HI temperature maps were produced in the frequency range $400 - 800\,{\rm MHz}$
        ($0.78\lesssim z \lesssim 2.55$) at intervals of $\delta\nu\sim1\,{\rm MHz}$, 
        corresponding to radial separations of $r_{\parallel}\sim 5-6 \,{\rm Mpc}/h$.
        These maps were created using the HEALPix pixelisation scheme \citep{2005ApJ...622..759G}
        with a resolution parameter {\tt nside}=512 ($\delta\theta\sim0.115^\circ$), corresponding to
        transverse scales $3.8\,{\rm Mpc}/h<r_{\perp}<8.2{\rm Mpc}/h$.
        
      \paragraph*{The foregrounds.}
        The foregrounds included in these simulations can be classified into two
        categories: isotropic and anisotropic. We expect that extra-galactic foregrounds,
        due to astrophysical sources beyond our galaxy should be isotropically
        distributed across the sky, while galactic foregrounds should be significantly
        more prominent closer to the galactic plane.
        
        The most relevant foreground in this frequency range in terms of amplitude is
        unpolarised galactic synchrotron emission. In broad terms, synchrotron emission was
        simulated by extrapolating the $408\,{\rm MHz}$ Haslam map \citep{1982A&AS...47....1H} to
        other frequencies using a direction-dependent spectral index provided by the Planck Sky
        model \citep{2013A&A...553A..96D}. Structure on angular scales smaller than the ones
        resolved in the original Haslam map and a mild frequency decorrelation was included by
        adding a constrained Gaussian realisation with parameters given by
        \citep{2005ApJ...625..575S}. Further details can be found in \citet{2014arXiv1405.1751A}.
        
        Other relevant foregrounds are point sources and free-free emission. We have
        modelled these as Gaussian realisations of the generic power spectra:
        \begin{equation}\label{eq:sck}
          C_l(\nu_1,\nu_2)=A\,\left(\frac{l_{\rm ref}}{l}\right)^{\beta}
          \left(\frac{\nu_{\rm ref}^2}{\nu_1\nu_2}\right)^{\alpha}
          \exp\left(-\frac{\log^2(\nu_1/\nu_2)}{2\xi^2}\right),
        \end{equation}
        with parameters given by \citet{2005ApJ...625..575S} (SCK from here on; see also Table 1
        in \citet{2014arXiv1405.1751A}). As mentioned in section \ref{ssec:ica}, ICA and PCA
        will only yield different results for non-Gaussian foregrounds, and therefore we
        will not be able to observe the advantage of using ICA over PCA for point sources
        and free-free emission. Any difference found between both methods will then be
        due only to the non-Gaussian galactic synchrotron foreground.
        
        As has was mentioned in section \ref{sec:imap}, even though we expect the 21 cm
        signal to be largely unpolarised, the presence of polarised foregrounds should be a
        concern for intensity mapping experiments due to polarisation leakage. Since our aim
        in this paper is to compare the usefulness of different foreground cleaning methods,
        and since the amplitude of polarised foregrounds depends both on instrumental design
        and survey strategy, we have decided to postpone the analysis of polarised foreground
        removal for future work. Nevertheless, in section \ref{ssec:corrlength} we have studied
        the minimum degree of frequency correlation that foregrounds must have in order to be
        efficiently removed by blind methods.
        
      \paragraph*{Instrumental effects.}
        \begin{table}
          \begin{center}
            \begin{tabular}{l|c}
              \hline
              $D_{\rm dish}$          & 15 m\\
              $t_{\rm obs}$ & 10000 h\\
              $\delta\nu$ & 1 MHz\\
              $N_{\rm dish}$ & 254\\
              $T_{\rm inst}$ & 25 K\\
              $(\nu_{\rm min},\nu_{\rm max})$ & $(400,800)$ MHz\\
              $f_{\rm sky}$ & 0.5\\
              \hline
            \end{tabular}
          \end{center}
          \caption{Instrumental parameters used for the simulations.}
                 \label{tab:inst}
        \end{table}
        The final observed temperature maps were generated by including the most relevant
        instrumental effects corresponding to a single-dish experiment (i.e. an instrument
        made up of a number of single-dish antennae used as a set of auto-correlators) with
        the parameters listed in Table \ref{tab:inst}. These are similar to those planned for
        the SKA-MID configuration, for which a single-dish strategy is optimal for
        cosmological purposes \citep{2014arXiv1405.1452B}. Two instrumental effects were
        implemented: a frequency-dependent beam and uncorrelated instrumental noise.
        
        Ideally we can parametrise the beam of such a system as being Gaussian with a
        simple frequency-dependent width:
        \begin{equation}\label{eq:beam_fwhm}
          \theta_{\rm FWHM}\sim\frac{\lambda}{D_{\rm dish}},         
        \end{equation}
        where $D_{\rm dish}$ is the antenna diameter. For the frequency range under
        consideration this yields an angular resolution that ranges between $0.7^\circ$
        and $1.4^\circ$.
        
        Furthermore, we have described the instrumental noise as being Gaussian
        and uncorrelated, with a frequency-dependent, per-pointing rms
        \begin{equation}\label{eq:sigma_noise}
          \sigma_N=T_{\rm inst}\,\sqrt{\frac{4\pi\,f_{\rm sky}}
          {\delta\nu\,t_{\rm obs}\,N_{\rm dish}\,\Omega_{\rm beam}}},
        \end{equation}
        where $T_{\rm inst}$ is the instrument temperature, $f_{\rm sky}$ is the observed
        sky fraction, $\delta{\nu}$ is the frequency resolution, $t_{\rm obs}$ is the
        total observation time, $N_{\rm dish}$ is number of dishes and
        $\Omega_{\rm beam}=1.133\,\theta_{\rm FWHM}^2$ is the beam solid angle. For the
        values listed in Table \ref{tab:inst} $\sigma_N$ varies between 0.025 and 0.05 mK.

        The process to build the full observed temperature maps is as follows:
        \begin{enumerate}
          \item We directly add the temperature maps corresponding to the cosmological
                signal and the different foregrounds. For computational ease and to
                avoid redundant calculations we then degrade the angular resolution
                of the total map from ${\tt nside}=512$ to 128, which is still
                perfectly compatible with the beam size ($\theta_{\rm pix}\sim0.45^\circ$).
          \item We smooth the total temperature using a Gaussian beam with
                full-width-half-maximum $\theta_{\rm FWHM}$.
          \item We add a random realisation of uncorrelated Gaussian noise with
                variance $\sigma_N$. Note that the per-pixel variance is similar to
                the one given in Equation \ref{eq:sigma_noise}, with the total
                number of pixels $N_{\theta}$ substituting the factor $4\pi\,
                f_{\rm sky}/\Omega_{\rm beam}$.
        \end{enumerate}
        At this point the observed maps should be cut to the desired survey mask.
        However, in order to simplify the analysis and to avoid the complication of
        accurately estimating the angular power spectrum of cut-sky maps, we
        decided to use full-sky maps for our fiducial simulations. The most important
        effect of an incomplete mask for foreground subtraction is a reduction in
        the number of pixels that can be used to determine the statistical properties
        of the foregrounds (e.g. estimating the covariance matrix for PCA or the
        negentropy for ICA). We have studied the relevance of this effect in section
        \ref{sssec:angmask}.
    
        Finally, we note that the parameters listed in Table \ref{tab:inst} correspond
        to our fiducial set of simulations. However, in order to study the effect of
        different instrumental configurations we have varied these parameters from
        their fiducial values. These variations are clearly stated in the corresponding
        sections below.
    
    \subsection{Clustering analysis}\label{ssec:analysis}
      \begin{table}
        \begin{center}
          \begin{tabular}{|c|c|c|c|c|}
            \hline
            \# & $\Delta\nu$ (MHz) & $z$ range & $\Delta\chi$ &
             $k_{\parallel,{\rm max}}$\\
            \hline
            1 & 436-537 & 1.64 - 2.25 & $636\,{\rm Mpc}/h$ & $0.50\,h\,{\rm Mpc}^{-1}$\\
            2 & 537-638 & 1.22 - 1.64 & $565\,{\rm Mpc}/h$ & $0.57\,h\,{\rm Mpc}^{-1}$\\
            3 & 638-739 & 0.92 - 1.22 & $491\,{\rm Mpc}/h$ & $0.64\,h\,{\rm Mpc}^{-1}$\\
            \hline
            \end{tabular}
          \end{center}
          \caption{Frequency bins used for the analysis of the radial power spectrum.
                   $\Delta\chi$ is the radial comoving distance between the
                   bin edges, and $k_{\parallel,{\rm max}}$ is the maximum radial
                   wavenumber reached in that bin given the frequency resolution. For the
                   three bins the minimum radial wavenumber $k_{\parallel,{\rm min}}
                   \equiv2\pi/\Delta\chi$ is approximately $0.01\,h\,{\rm Mpc}^{-1}$.
                   Note that these bins were only used for the analysis of the radial
                   power spectrum, while the foreground cleaning was done on the full
                   frequency band ($400\,{\rm MHz}<\nu<800\,{\rm MHz}$).}
                   \label{tab:radbins}
        \end{table}
      The most important cosmological constraints from HI intensity mapping will probably
      be obtained from the two-point statistics of the overdensity field, since this
      encapsulates the vast majority of the information regarding the underlying cosmological
      model. Hence the merit of a given foreground cleaning method should be judged partly in
      terms of its ability to recover the true power spectrum on different radial and angular
      scales. We have therefore defined three quantities to characterise the efficiency
      of the different cleaning methods:
      \begin{gather}\nonumber
        \eta\equiv\frac{\langle P_{\rm clean}-P_{\rm true}\rangle}{\sigma_P},\,\hspace{12pt}
        \rho\equiv\frac{\langle P_{\rm res}\rangle}{\sigma_P},\\\label{eq:fom}
        \epsilon\equiv\left\langle\frac{P_{\rm clean}-P_{\rm true}}{P_{\rm true}}\right\rangle,
      \end{gather}
      where $P_{\rm clean},\,P_{\rm true},\,P_{\rm res}$ are the power-spectra of the
      cleaned maps, the true signal (cosmological signal and noise) and the residuals
      (``${\rm cleaned}-{\rm true}$'') respectively, and $\sigma_P$ is the statistical error in
      the power-spectrum, caused both by clustering variance and instrumental noise. Thus
      $\epsilon$ and $\eta$ quantify the bias induced by foreground subtraction on the power
      spectrum as a fraction of the true power spectrum and of the expected statistical
      uncertainties respectively, while $\rho$ is a measure of the corresponding loss in the
      signal itself. A perfect foreground cleaning would yield 0 for these three quantities.
      
      We must clarify that we have used generic term ``power spectrum'' here referring to both the
      angular power spectrum $C_l$ and the radial power spectrum (defined below), which
      depend on $(\nu,l)$ and $(z_{\rm eff},k_{\parallel})$ respectively. This
      distinction is important: since the foregrounds are assumed to be extremely smooth
      in frequency (i.e. the radial direction) we can expect the effects of foreground
      subtraction to be reduced to the largest radial scales (an effect commonly known as
      the ``foreground wedge'' \citep{2014PhRvD..90b3018L}).
      
      It is also important to note that we have computed the ensemble averages in
      Equation \ref{eq:fom} by averaging over the 100 different realisations of the cosmological
      signal, and the foregrounds. We believe that, given the current uncertainties in the 
      multi-frequency description of the different relevant foregrounds, this approach is
      reasonably conservative in that, by averaging over foreground realisations, we effectively
      marginalise over these uncertainties. Consequently, the mean and variance of the quantities
      in Equation \ref{eq:fom} will depend not only on the statistics of the cosmological signal,
      but also on the model used to describe the foregrounds.
      
      \subsubsection{Angular power spectrum.}
        Assuming a full-sky survey, for a fixed frequency shell, the angular power spectrum of
        the brightness temperature fluctuations $\Delta T$ is estimated by first computing their
        harmonic coefficients
        \begin{equation}
          a_{lm}(\nu)=\int d\nv^2\,\Delta T(\nu,\nv)Y^*_{lm}(\nv),
        \end{equation}
        where $Y_{lm}(\nv)$ are the spherical harmonics. The angular power spectrum
        $C_l\equiv\langle |a_{lm}|^2\rangle$ is then estimated by averaging over the symmetric
        index $m$
        \begin{equation}
          \tilde{C}_l\equiv\frac{1}{2l+1}\sum_{m=-l}^l|a_{lm}|^2.
        \end{equation}
      
        Unfortunately this estimator is biased and non-optimal for cut-sky maps, and in fact the
        estimation of angular power spectra for incomplete maps is a non-trivial problem that has
        been widely treated in the literature \citep{1997PhRvD..55.5895T,2001PhRvD..64h3003W,
        2002ApJ...567....2H}. As we mentioned above, our fiducial set of simulations 
        were generated
        as full-sky maps in order to avoid these complications. However we have studied the effect
        of an angular mask in section \ref{sssec:angmask}, and in this case the angular power
        spectra were computed using the {\tt PolSpice} software package
        \citep{2004MNRAS.350..914C}.
      
      \subsubsection{Radial power spectrum.}
        \begin{figure}
          \centering
          \includegraphics[width=0.49\textwidth]{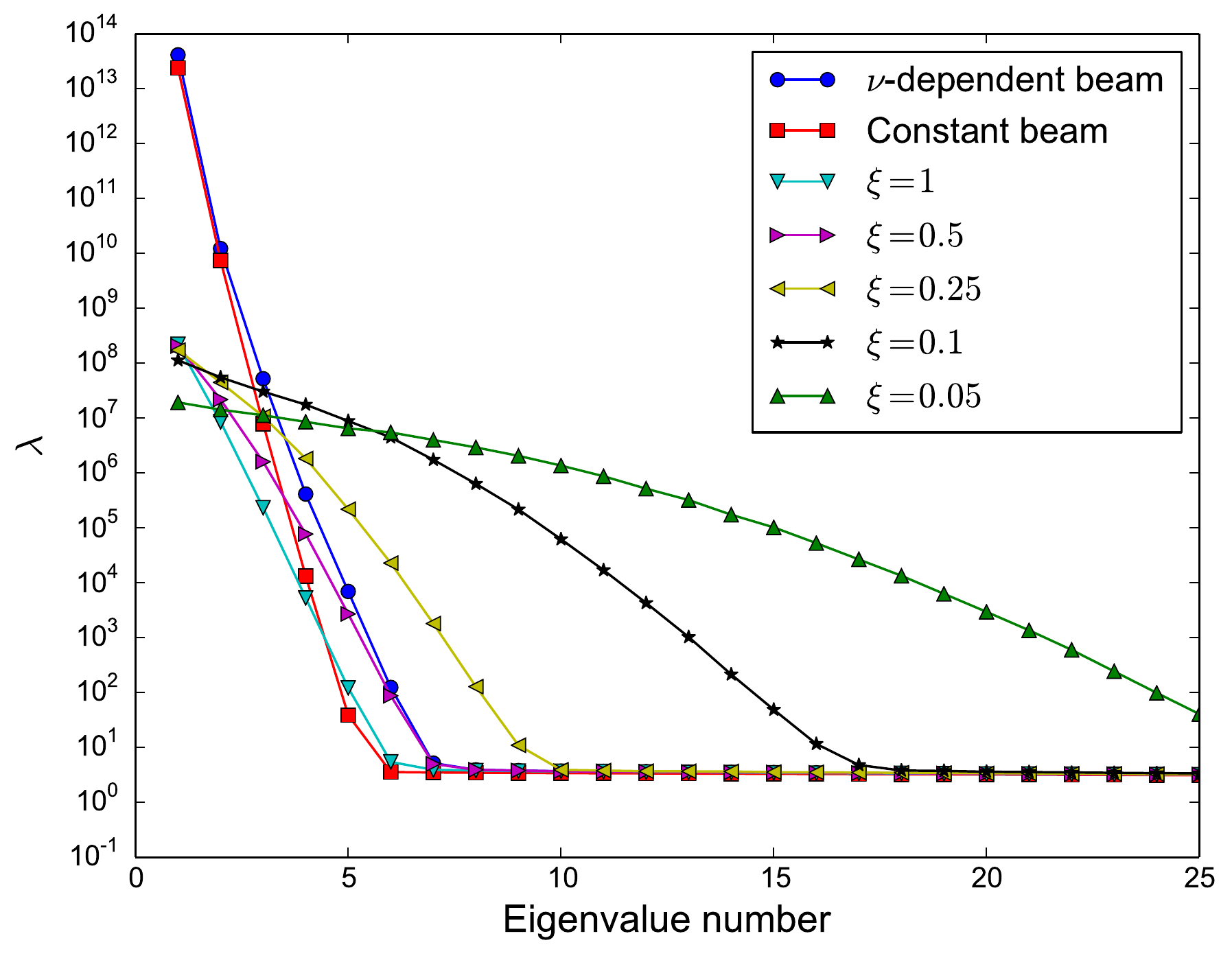}
          \caption{Principal eigenvalues of the frequency covariance matrix for a simulation
                   including a frequency-dependent beam (blue circles) and a constant beam
                   (red squares), as well as for simulations containing foregrounds with
                   different correlation lengths $\xi\in(1,0.05)$. In the first two cases
                   there is a clear division between foreground eigenvalues and cosmological
                   ones, although the presence of a $\nu$-dependent beam makes the transition
                   between the two smoother, and a larger number of foreground degrees of
                   freedom must be eliminated. For smaller frequency correlation lengths
                   the contribution of foregrounds spreads through a larger number of eigenvalues,
                   making foreground contamination more troublesome.}
                   \label{fig:eigvals}
        \end{figure}
    \begin{figure*}
      \centering
      \includegraphics[width=0.49\textwidth]{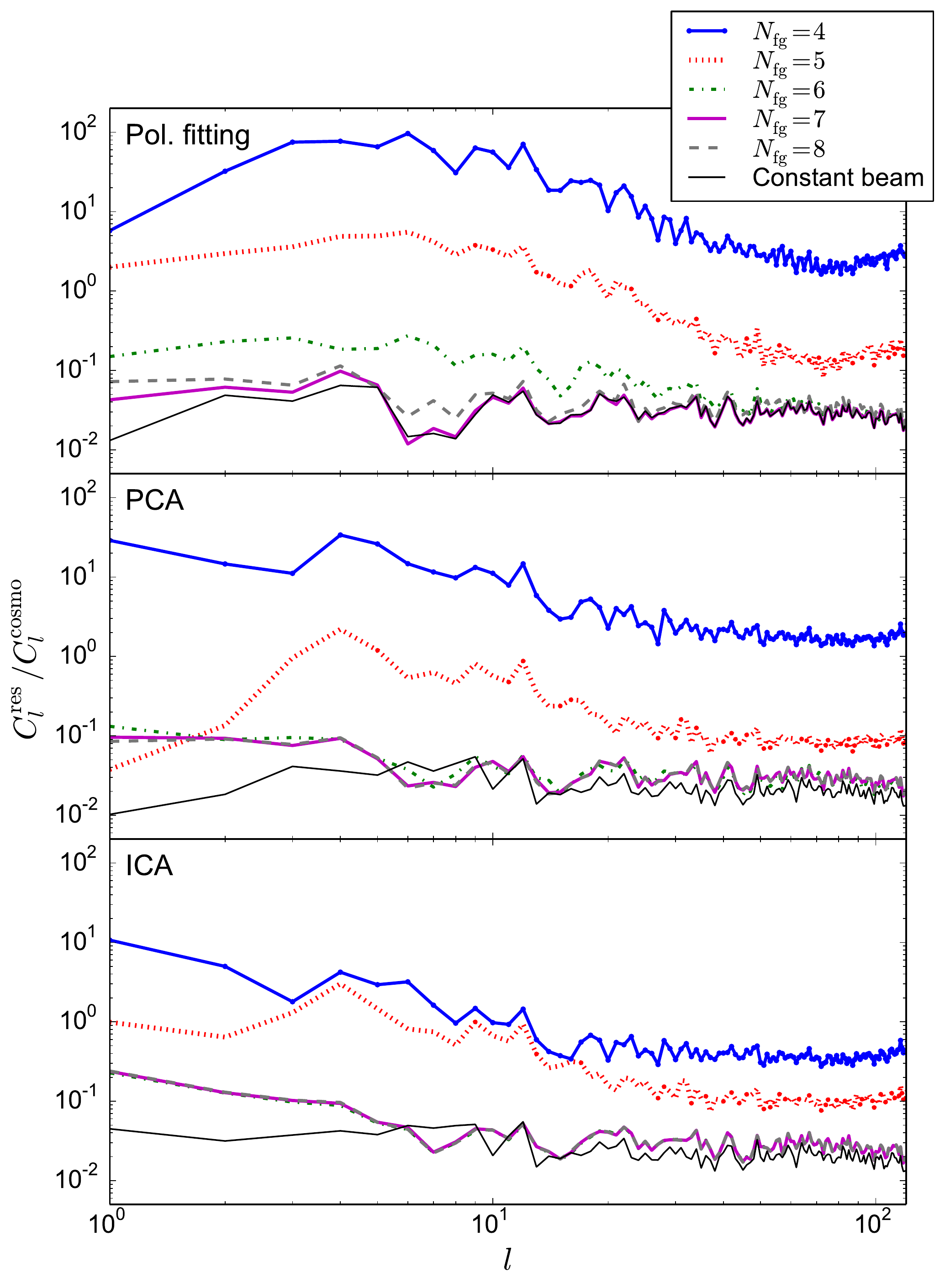}
      \includegraphics[width=0.49\textwidth]{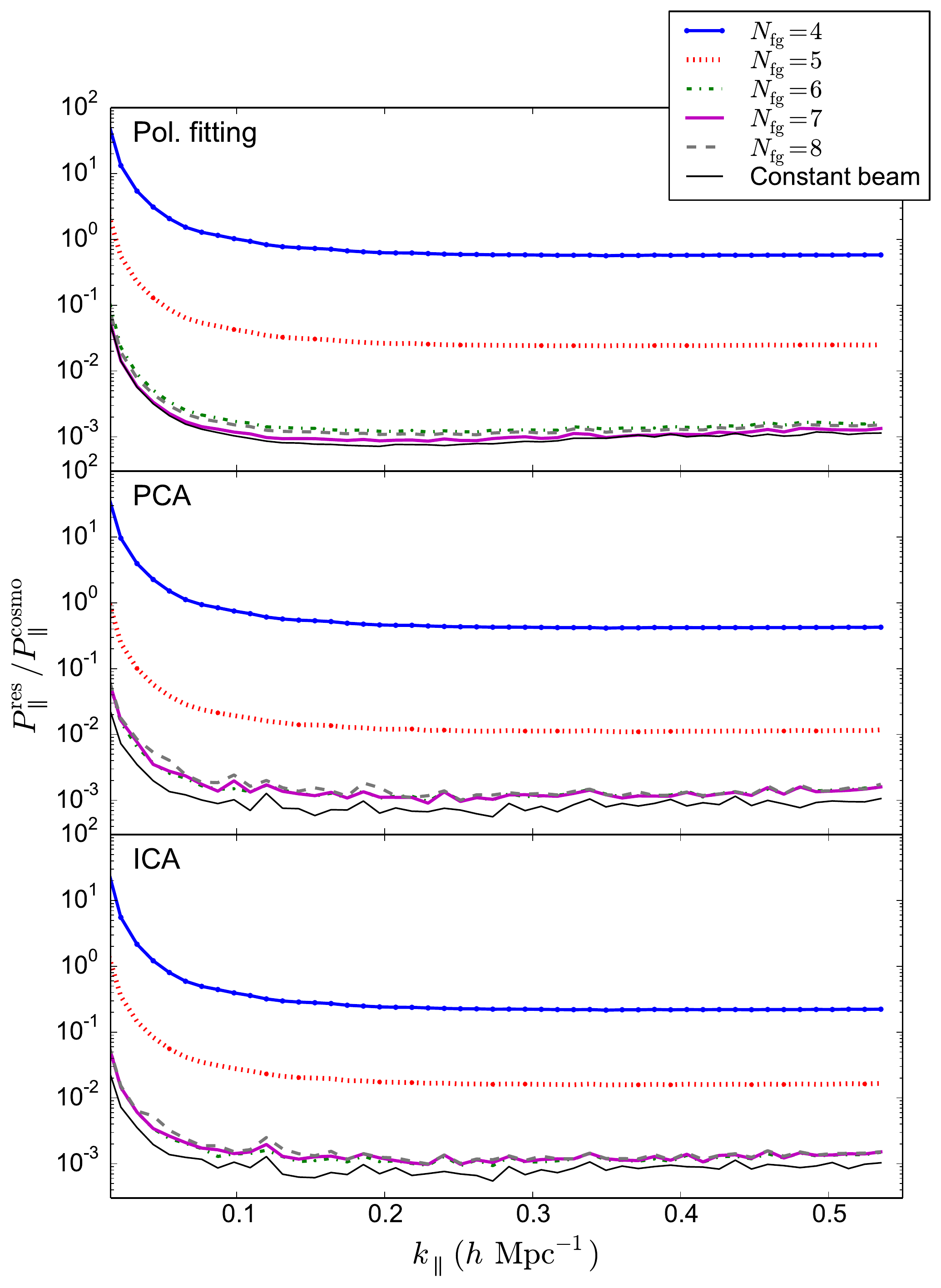}
      \caption{Ratio of the power spectrum of the foreground cleaning residuals to the
               power spectrum of the cosmological signal as a function of the number of
               foreground degrees of freedom removed for the three different cleaning methods.
               The left and right panels show the results for the angular and radial power
               spectrum respectively. Each plot contains 3 sub-panels, showing the results for
               polynomial fitting, PCA, and ICA (in descending order). The angular power spectra
               shown correspond to a frequency bin $599\,{\rm MHz}<\nu<600\,{\rm MHz}$, while the
               radial ones correspond to the second bin in Table \ref{tab:radbins}. As can be
               seen, in most cases the efficiency of the foreground cleaning converges for
               $N_{\rm fg}\sim 6-7$, although the distinction between both values is clearer for
               polynomial fitting, which attains a clear minimum for $N_{\rm fg}=7$. This result
               could have been anticipated visually from Figure \ref{fig:eigvals}. The black
               lines in both plots correspond to the optimal case for the
               constant-beam simulation ($N_{\rm fg}=7$ for polynomial fitting and $N_{\rm fg}=5$
               for PCA and ICA).} \label{fig:nfg}
    \end{figure*}
        Because astronomical observations are performed on the lightcone, 
        cosmological observables such as the temperature perturbation $\Delta T$ are 
        not homogeneous in the
        radial direction, since they evolve in time. For this reason, the only rigorous method to
        analyse the clustering pattern of the overdensity field in the radial direction is
        indirectly through the angular cross-correlation of maps at different redshifts
        \citep{2012PhRvD..86f3503M,2012MNRAS.427.1891A}. However, in practice it is possible to
        divide the survey into redshift shells that are wide enough to contain the most relevant
        radial scales and narrow enough so that we can assume a static universe at some effective
        redshift $z_{\rm eff}$, neglecting evolution effects. This has in fact been common
        practice among most galaxy redshift surveys \citep{2014MNRAS.441...24A,
        2013MNRAS.430..924C}, and enables much more direct, easier access to important
        cosmological observables. In order to obtain a direct intuition about the effectiveness
        of foreground subtraction in the radial direction we have followed this approach here.

        We have divided our full frequency range into a number of thick redshift shells of equal
        bandwidth $\Delta\nu\sim100\,{\rm MHz}$, each containing the same number of frequency bins
        (see Table \ref{tab:radbins} for details). The location, width and number of these shells
        were chosen in order to avoid the edge effects described in section \ref{sssec:edges} and
        to include the most relevant cosmological scales. For each of these shells, every
        individual pixel corresponds to a different realisation of the overdensity field along the
        line of sight. We can therefore compute the radial power spectrum by averaging over the
        modulus of the Fourier transform of each line of sight
        \begin{equation}
          P_{\parallel}(k_{\parallel})\equiv\frac{\Delta\chi}{2\pi N_\theta}\sum_{i=1}^{N_\theta}
          |\widetilde{\Delta T}(\hat{\bf n},k_{\parallel})|^2,
        \end{equation}
        where $\Delta\chi=\chi(z_{\rm max})-\chi(z_{\rm min})$ is the comoving width of the
        redshift shell under consideration. We estimate the Fourier coefficients
        $\widetilde{\Delta T}$ by computing the Fast Fourier Transform (FFT) of each line of
        sight. Since the different values of $\Delta T$ are separated by a constant
        frequency interval $\delta \nu=1\,{\rm MHz}$, its FFT is calculated at constant
        interval in the conjugate variable corresponding to $\nu$, $\delta k_\nu=2\pi/\Delta \nu$.
        Assuming a constant effective redshift $z_{\rm eff}$ for each bin, $k_\nu$ can be related
        to the radial wavenumber $k_\parallel$ through
        \begin{equation}
          k_\parallel=\frac{\nu_{21}H(z_{\rm eff})}{(1+z_{\rm eff})^2}\,k_\nu.
        \end{equation}
        A more rigorous definition for $P_{\parallel}(k_{\parallel})$ can be found in
        \citet{2014arXiv1405.1751A}.

  \section{Results}\label{sec:results}
    We have used 100 independent simulations as described in the previous section in order
    to compare the three blind subtraction methods and evaluate the ranges of scales in which
    the recovered maps can be reliably used for cosmology. We have studied the number of
    foreground degrees of freedom $N_{\rm fg}$ that need to be removed (i.e. the dimension of
    ${\bf s}$ in Equation \ref{eq:bfg_eq}) for each method, the values of the parameters defined
    in Equation \ref{eq:fom} as a function of scale and method, and the limits of applicability
    of these techniques.\footnote{We have made available a public version of the computer codes
    used to carry out this analysis at \url{http://intensitymapping.physics.ox.ac.uk/codes.html}.}
      
    \subsection{Foreground degrees of freedom}\label{ssec:nfg}
      By subtracting the foregrounds we are effectively removing information from the temperature
      maps. We should therefore be concerned about removing only the information related to the
      foregrounds, and minimising the loss of cosmological information. This implies minimising
      the number of degrees of freedom that we subtract, which ideally would correspond to the
      number of independent foregrounds.
      
      In our case we have included four different foregrounds (galactic synchrotron, point
      sources and galactic and extra-galactic free-free). Additionally, the mean HI 
      temperature is also a smooth function of the frequency, and will be inevitably removed
      by the foreground cleaning method. On top of this, a frequency-dependent instrumental
      beam will also act as an extra effective foreground, yielding a total of $\sim6$ potential
      foreground sources, the cleaning of which will depend critically on them being clearly
      distinct from the cosmological signal we hope to measure.
      
      PCA offers an intuitive way to address this question, and to decide on the number
      of foreground degrees of freedom to subtract. We know that most of the information related
      to the foregrounds should be contained in a few of the largest eigenvalues of the frequency
      covariance matrix. Thus, by looking at these eigenvalues it is possible to see whether 
      a number of them is clearly different (much larger) from the rest, and whether there exists
      a clear divide between foreground and cosmological eigenvalues. Figure \ref{fig:eigvals}
      shows the first 25 principal eigenvalues for several of our simulations. One of them (blue
      circles) includes a frequency-dependent beam size given by Eq. (\ref{eq:beam_fwhm}), while
      another one (red squares) was generated for a constant beam size corresponding to the median
      frequency of the simulation. It is possible to see that in both cases there exists a clear
      distinction between foreground and cosmological eigenvalues, although the presence of a 
      frequency-dependent beam makes the transition between the two smoother. We can predict an
      optimal value of $N_{\rm fg}=5$ for the constant-beam simulation, and either 6 or 7 for
      the $\nu$-dependent beam using PCA.
      
      We have studied the total number of eigenvalues to subtract by computing the ratio of
      the power spectrum of the residual maps to the cosmological power spectrum in one of
      our simulations for the three methods and for different numbers of foregrounds. Figure
      \ref{fig:nfg} shows this ratio for the angular (left panel) and radial (right panel) power
      spectra and for the three different methods (in descending order: polynomial fitting, PCA
      and ICA). For this plot we chose to show the results corresponding to an intermediate
      frequency bin $599\,{\rm MHz}<\nu<600\,{\rm MHz}$ for the angular case, and the second
      bin in Table \ref{tab:radbins} for the radial case, but similar results hold in general.
      As anticipated above, both PCA and ICA converge for $N_{\rm fg}=6$ or 7, and we do
      not gain anything by subtracting additional degrees of freedom. For polynomial fitting,
      however, the optimal value is clearly $N_{\rm fg}=7$. In view of this result we have
      performed the analysis below on the total ensemble of simulations using the fiducial
      value $N_{\rm fg}=7$ for the three methods.
      
    \subsection{The effects of foreground removal}\label{ssec:main_result}
     \begin{figure*}
       \centering
      \includegraphics[width=0.49\textwidth]{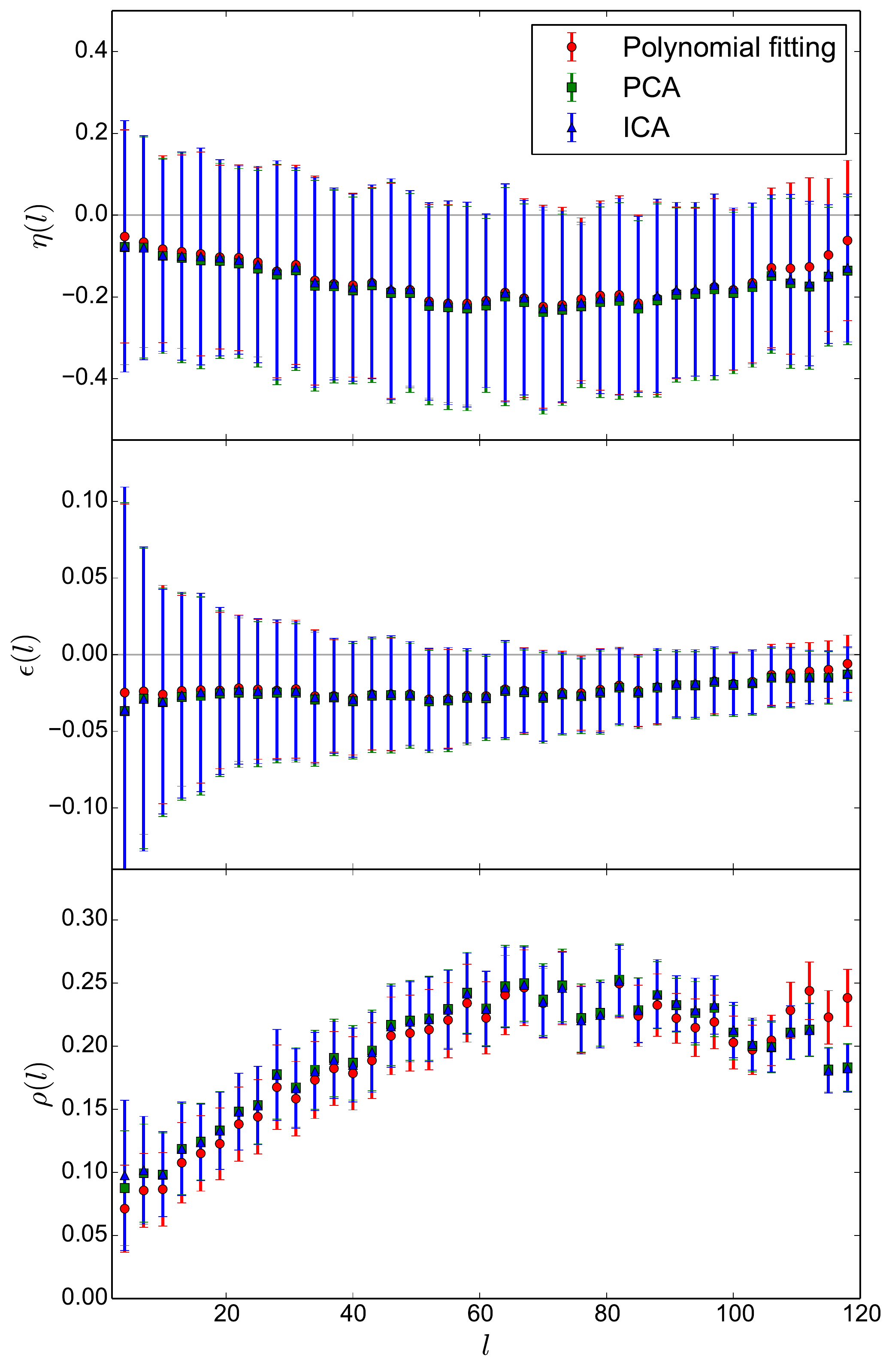}
      \includegraphics[width=0.49\textwidth]{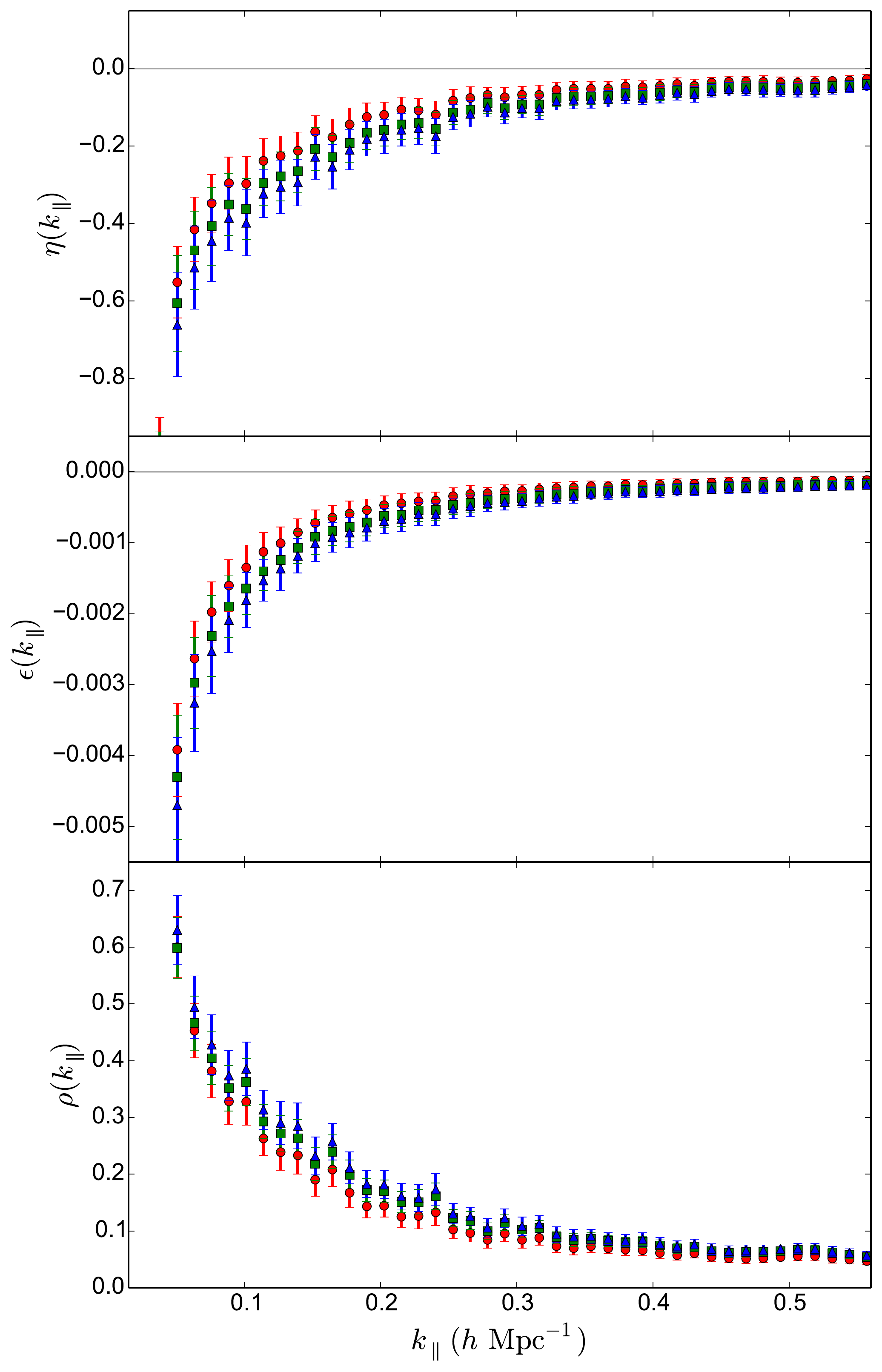}
       \caption{The parameters $\eta$, $\epsilon$ and $\rho$ (rows 1 to 3 respectively),
                introduced in Equation \ref{eq:fom}, for the angular power spectrum
                in the bin $\nu\in(599,600)\,{\rm MHz}$ (left panel) and the radial
                power spectrum for the second bin in Table \ref{tab:radbins} (right panel).
                Each plot shows the result for polynomial fitting (red circles), PCA (green
                squares) and ICA (blue triangles). The error bars in the plots show the
                variance of each of these quantities.}
       \label{fig:params_allmethods}
    \end{figure*}
    \begin{figure*}
      \centering
      \includegraphics[width=0.51\textwidth]{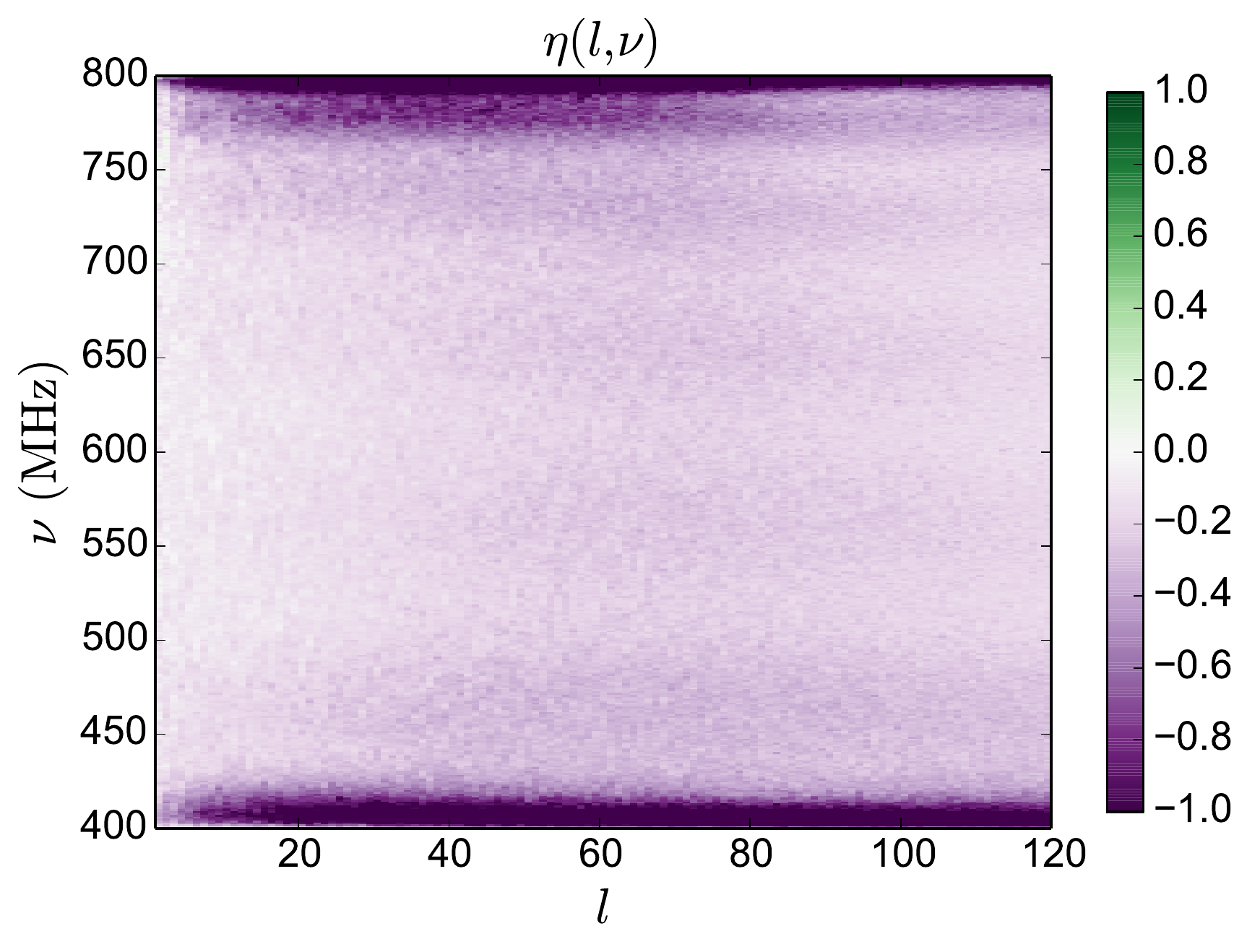}
      \includegraphics[width=0.47\textwidth]{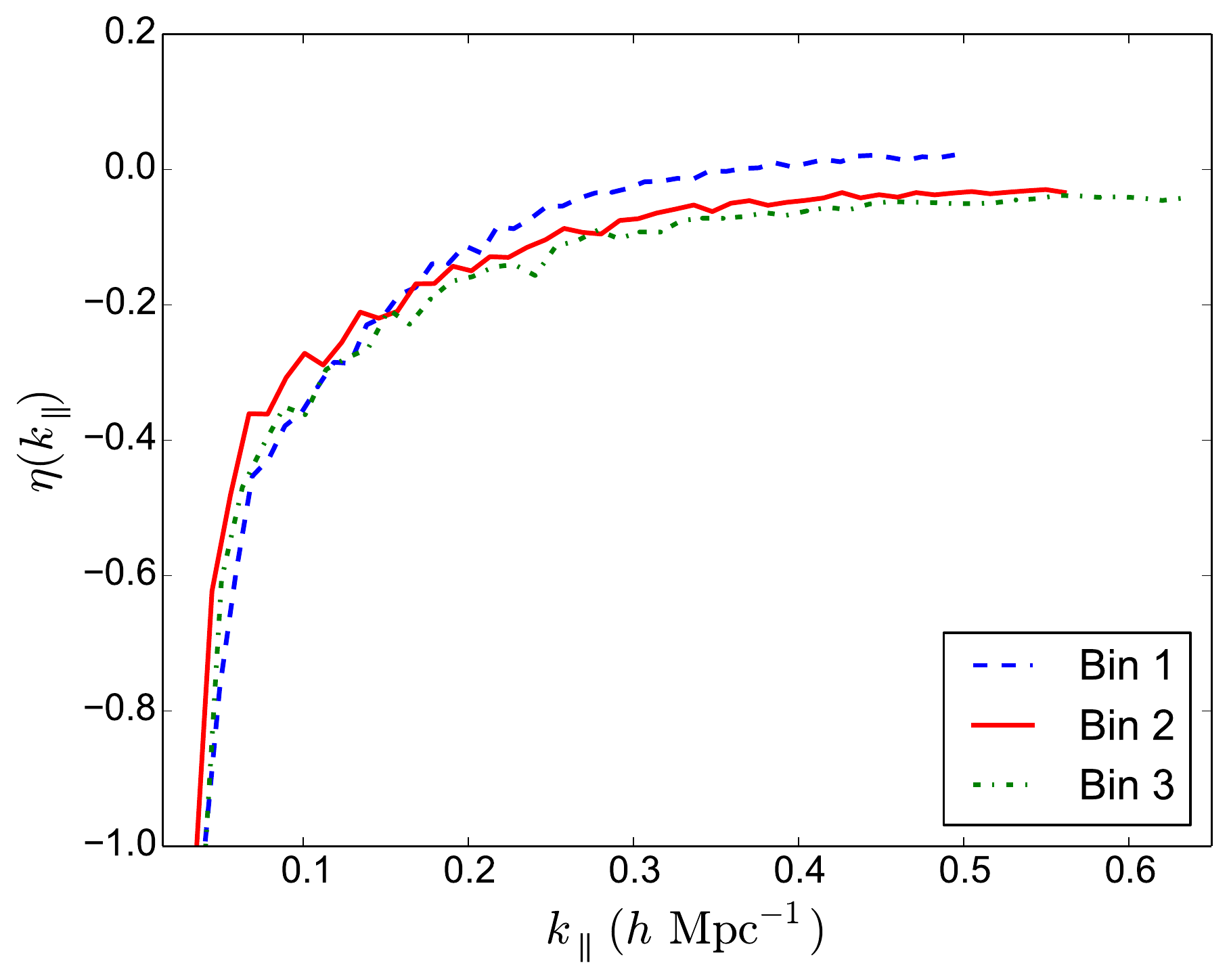}
      \includegraphics[width=0.51\textwidth]{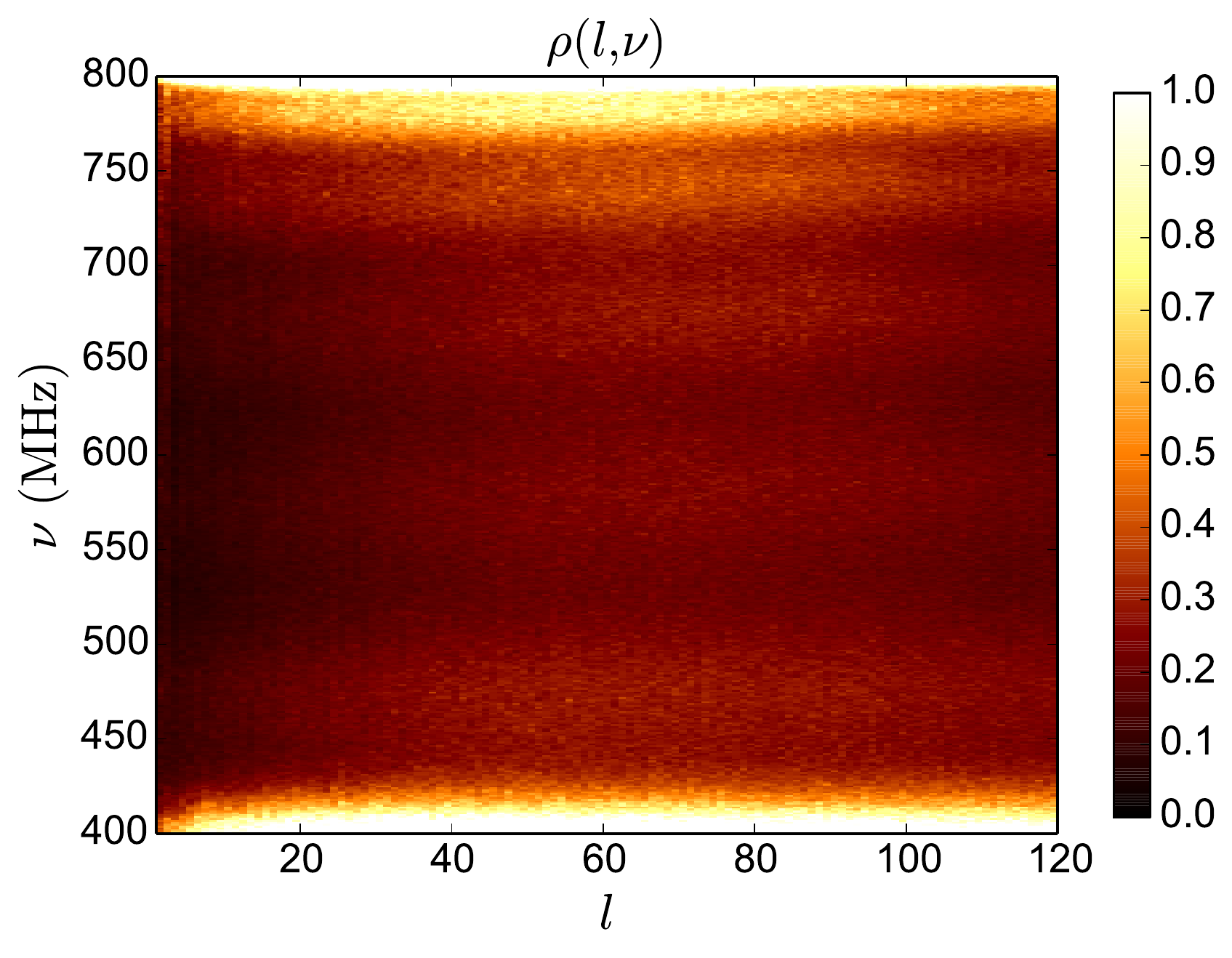}
      \includegraphics[width=0.47\textwidth]{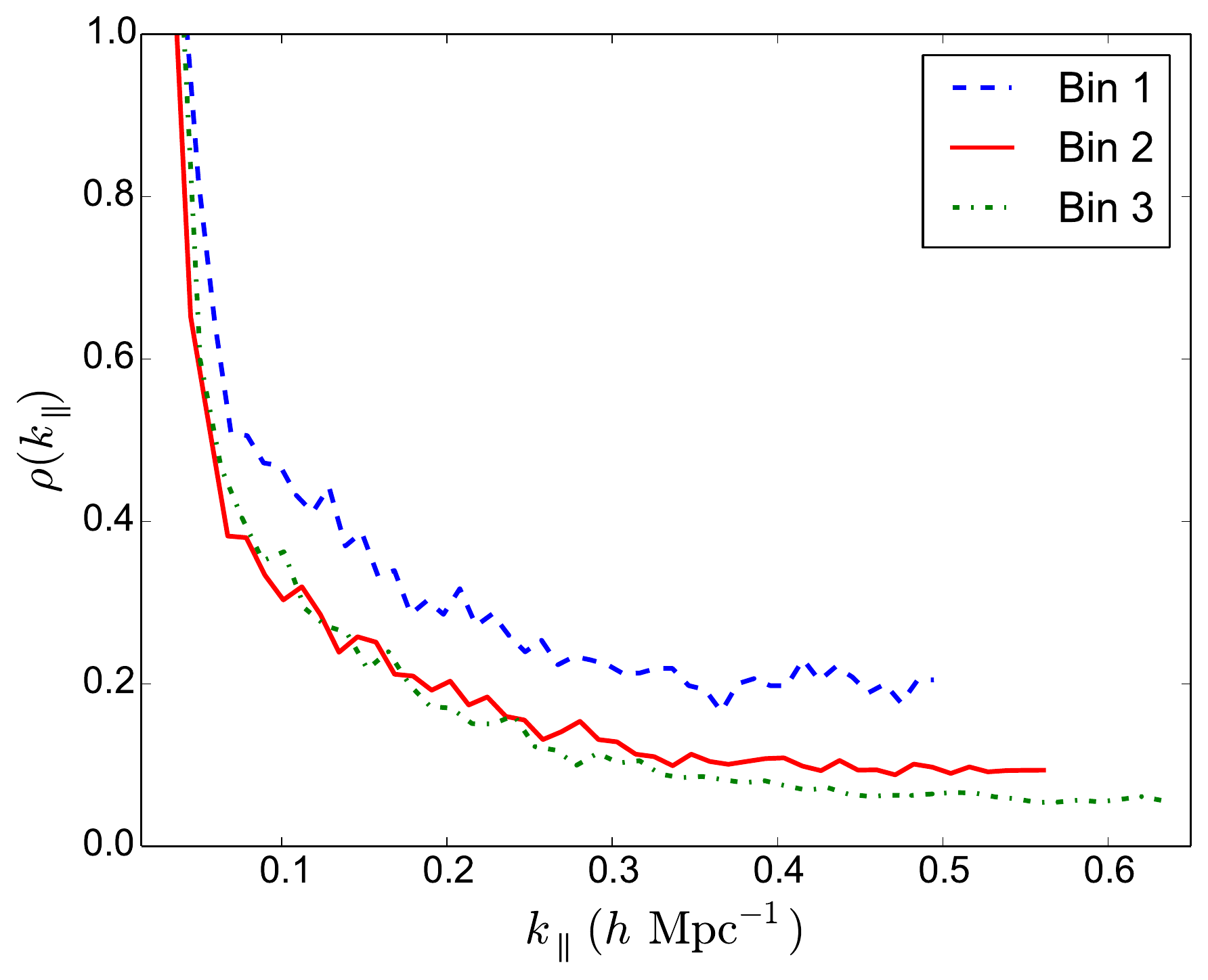}
      \caption{The parameters $\eta$ and $\rho$ from Eq. \ref{eq:fom} (top and bottom rows
               respectively) for the angular (left column) and radial (right column) power
               spectra for all the different frequency (or redshift) bins considered in this
               analysis. The results shown correspond to the PCA method, but equivalent 
               results were found for polynomial fitting and ICA. In most cases the bias due
               to foreground cleaning is much smaller than the statistical errors for a wide
               range of scales, although this breaks down on large radial scales and close
               to the edges of the frequency band.} \label{fig:params_pca_2d}
     \end{figure*}
      \begin{figure*}
        \centering
      \includegraphics[width=0.49\textwidth]{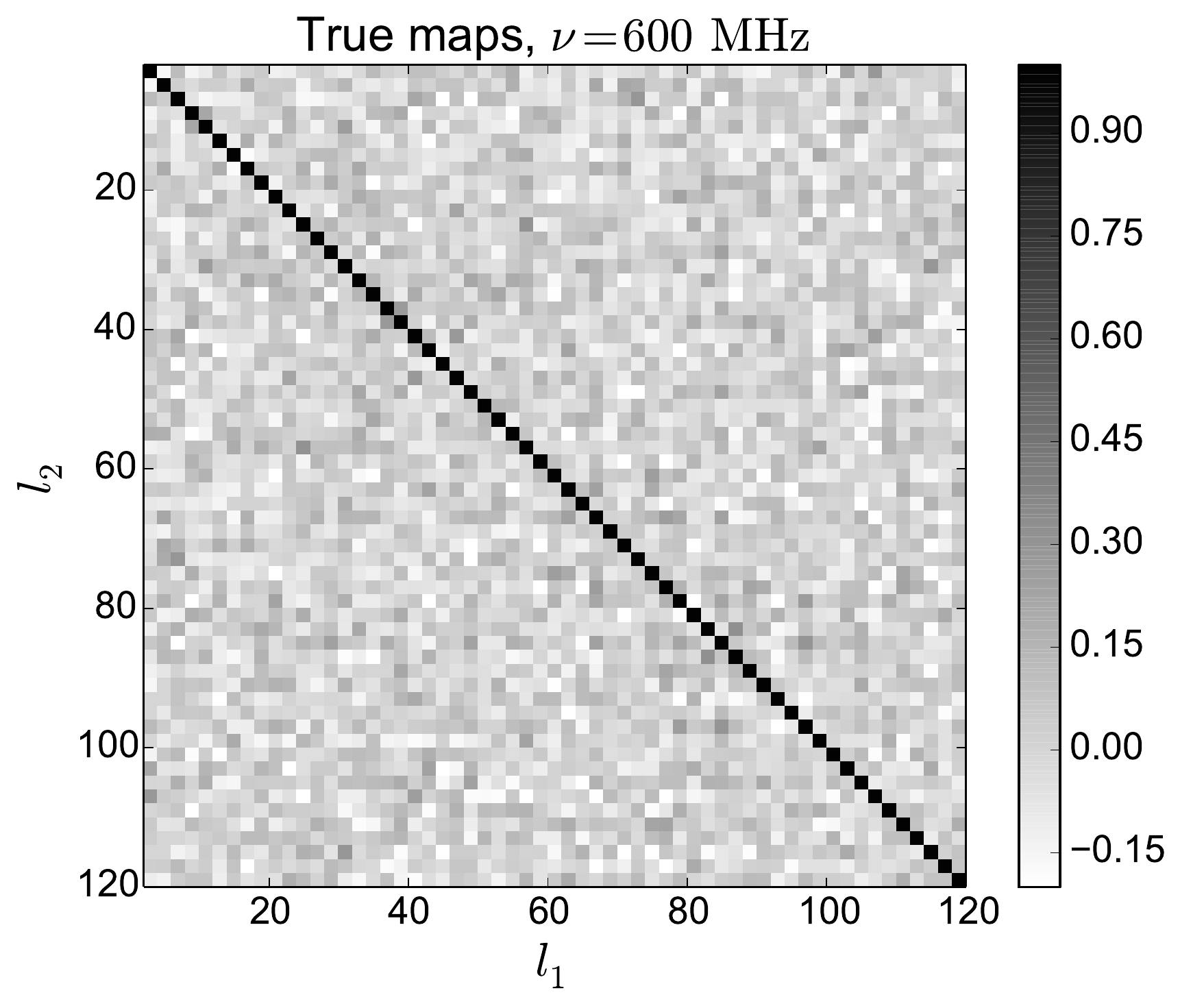}
      \includegraphics[width=0.49\textwidth]{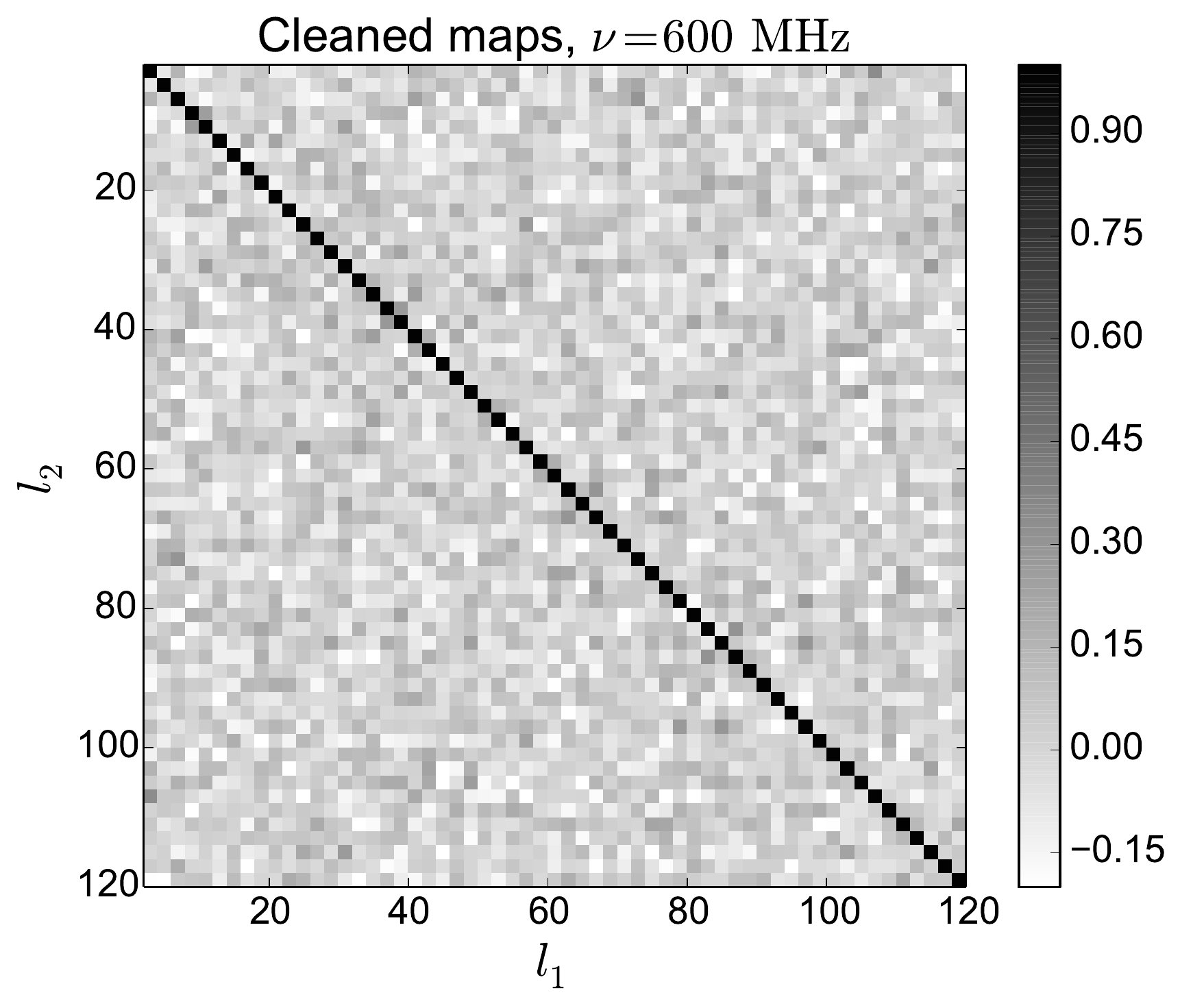}
      \includegraphics[width=0.49\textwidth]{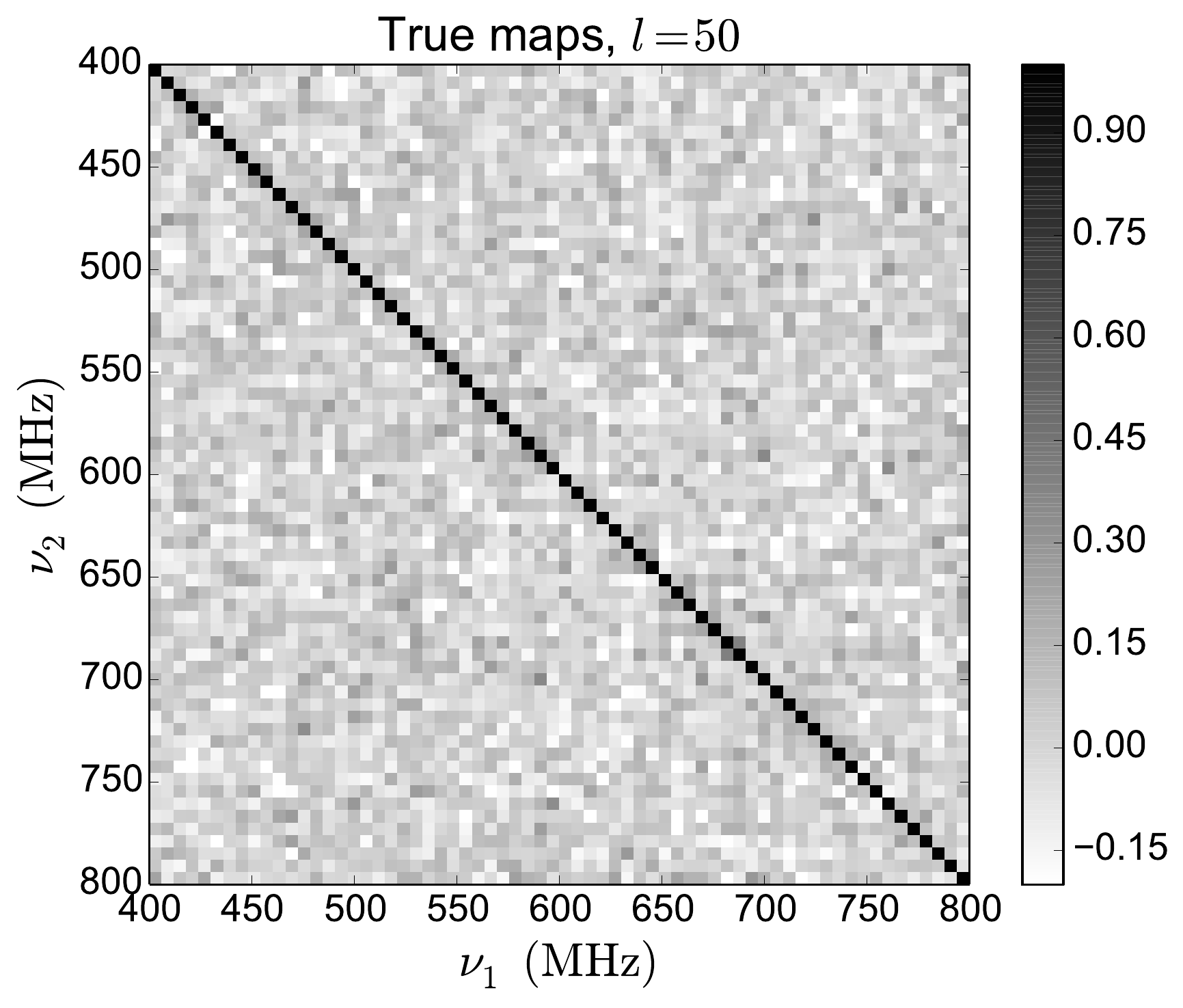}
      \includegraphics[width=0.49\textwidth]{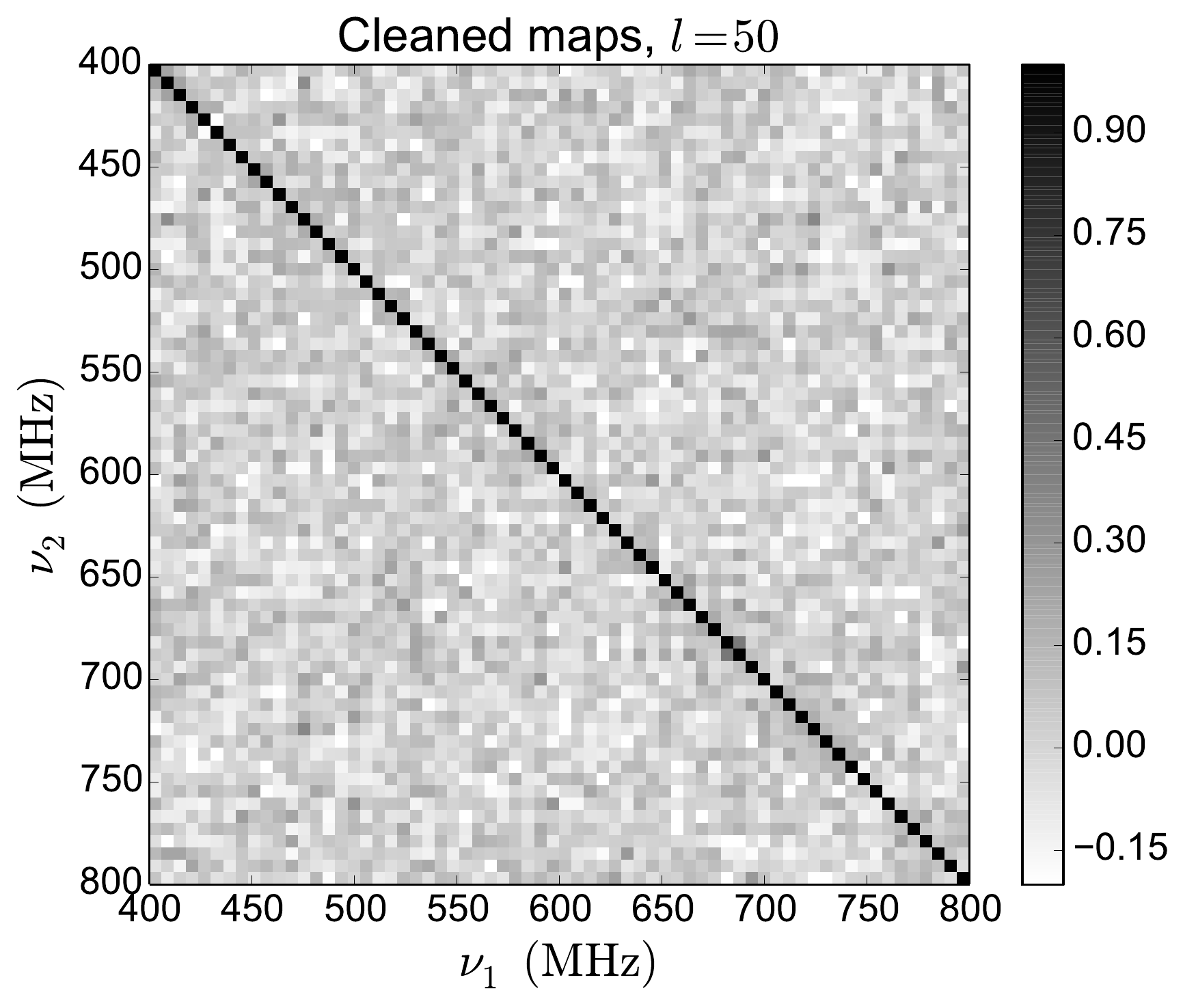}
        \caption{Correlation matrix of the angular power spectrum at a fixed frequency
                 $\nu=600\,{\rm MHz}$ (top row) and at a fixed scale $l=50$ (bottom row)
                 for the true signal (left column) and the cleaned maps (right column).
                 In order to reduce the statistical fluctuations the covariance matrices have been
                 rebinned in bins of $\Delta l=4$ and $\Delta \nu=6\,{\rm MHz}$.} \label{fig:corr}
      \end{figure*}
      \begin{figure*}
        \centering
      \includegraphics[width=0.49\textwidth]{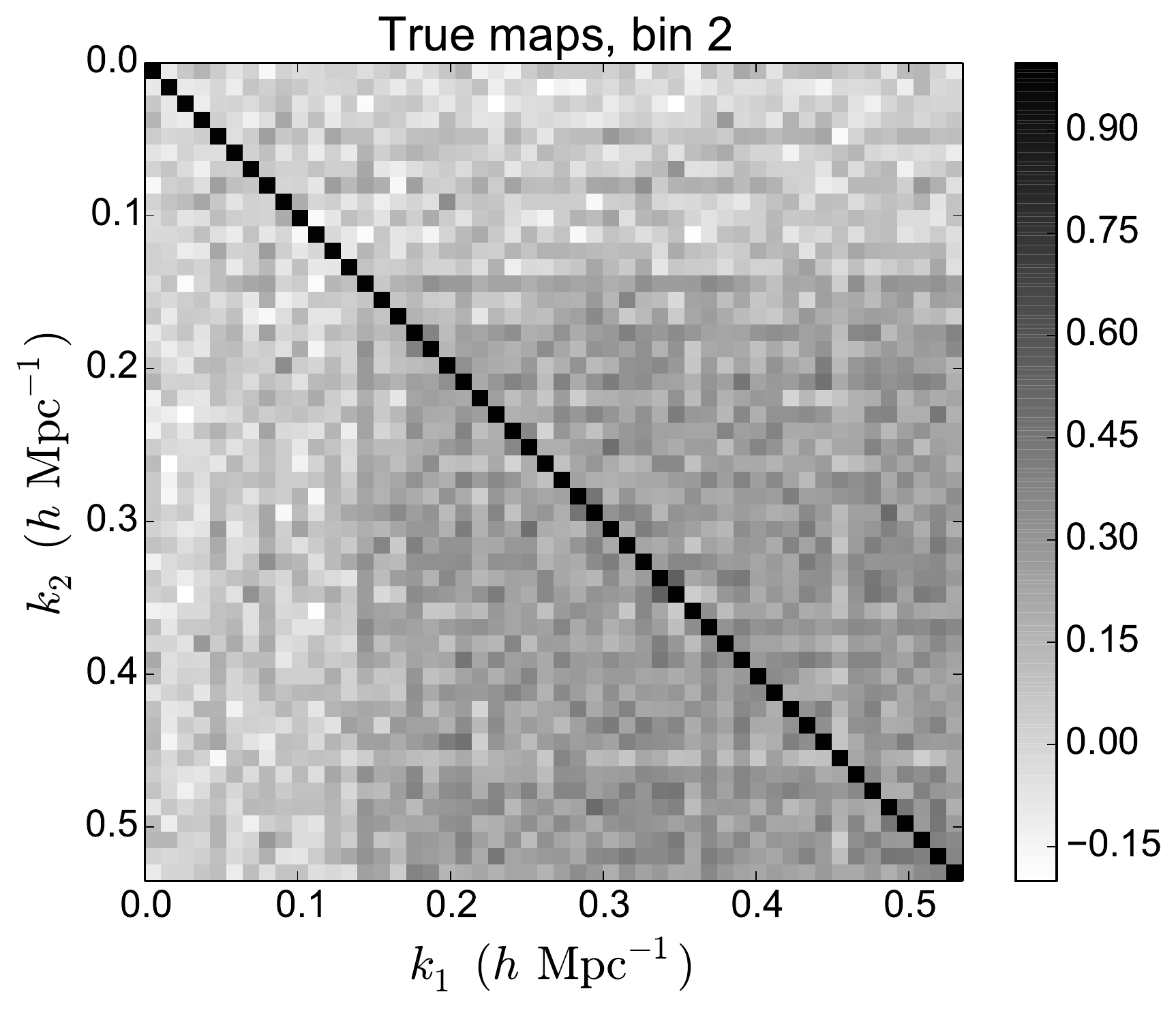}
      \includegraphics[width=0.49\textwidth]{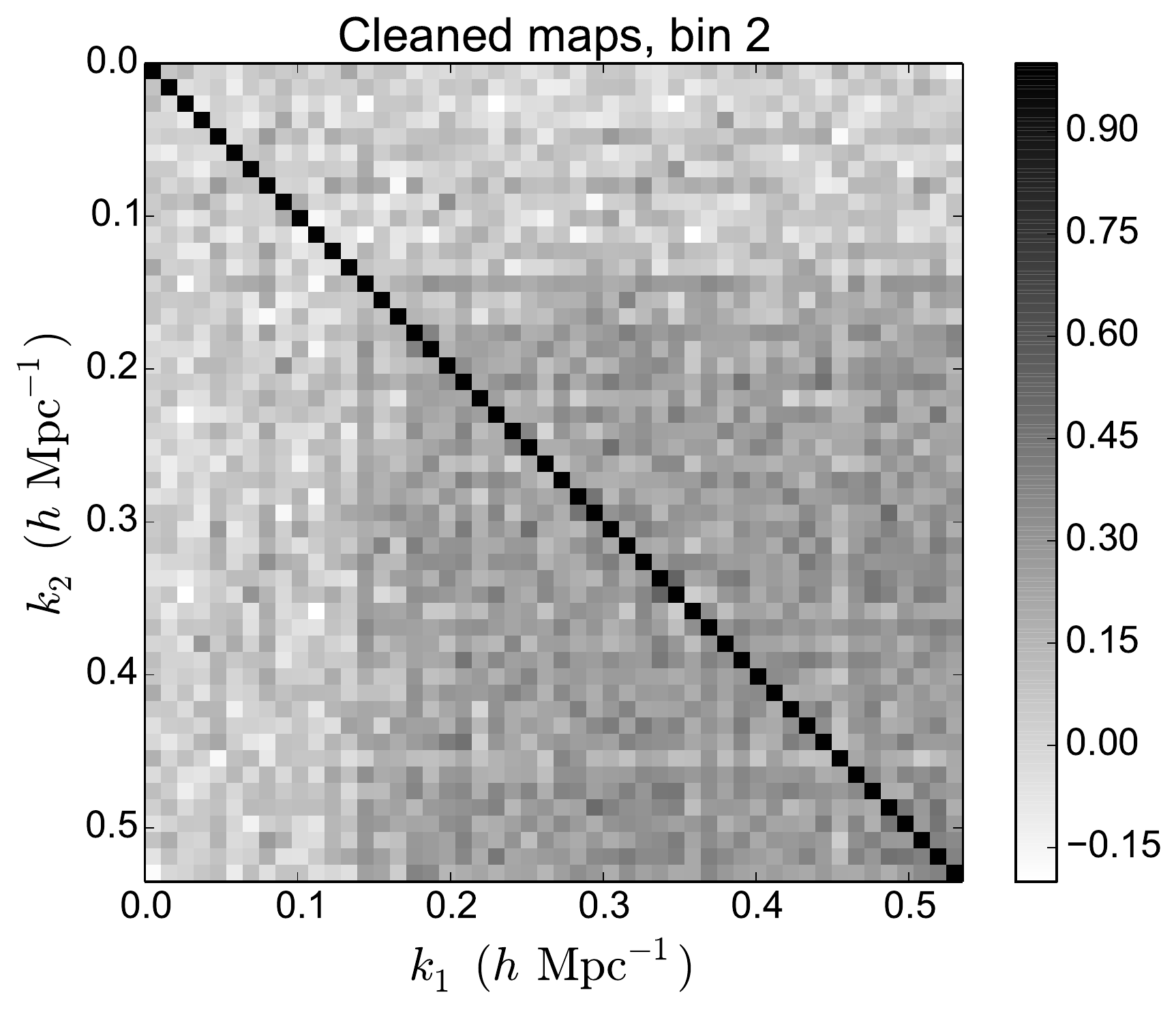}
        \caption{Correlation matrix of the radial power spectrum for the second bin in Table
                 \ref{tab:radbins} for the true cosmological signal (left panel) and the
                 cleaned maps (right panel).} \label{fig:corr_pkr}
      \end{figure*}
      In order to evaluate the extra statistical and systematic uncertainties introduced by
      the process of foreground subtraction, and to ascertain the range of scales where the
      cleaned maps can be reliably used to obtain cosmological constraints we have computed
      the parameters $\eta$, $\epsilon$ and $\rho$ introduced in Eq. (\ref{eq:fom}) from our
      100 independent simulations.
      
      Figure \ref{fig:params_allmethods} shows the values of these parameters computed for the
      angular power spectrum in the frequency bin $599\,{\rm MHz}<\nu<600\,{\rm MHz}$ (left 
      column) and for the radial power spectrum for the second bin in Table \ref{tab:radbins}
      (right column). Each plot shows the results for the three cleaning methods, and the error
      bars show the statistical deviation of these quantities. In all cases the three methods
      yield very similar (almost equivalent) results, all of them being able to clean the
      foregrounds reasonably well for the same number of foreground degrees of freedom. There
      is a very wide range of scales where foreground contamination is small in comparison with
      the expected statistical errors, and which can be reliably used for cosmology. In particular,
      PCA and ICA yield remarkably similar results, in spite of the latter exploiting the central
      limit theorem to enforce the statistical independence of the different foreground components.
      Thus, at least for the foregrounds simulated in this work, requiring the foregrounds to be
      uncorrelated (which is common to PCA and ICA) is the most relevant constraint. Furthermore,
      the results for polynomial fitting are very similar to the other two methods (even
      slightly better for certain radial scales), in spite of this method being the most na\"ive
      approach to foreground cleaning. It is also worth noting that both $\eta$ and $\rho$ become
      more consistent with 0 on large angular scales. This 	is due to the fact that these two
      quantities are weighed by the statistical errors, wich are inversely proportional to the
      square root of the number of modes available ($\sigma(C_l)\propto(2\,l+1)^{-1/2}$).
      
      \begin{figure*}
        \centering
      \includegraphics[width=0.49\textwidth]{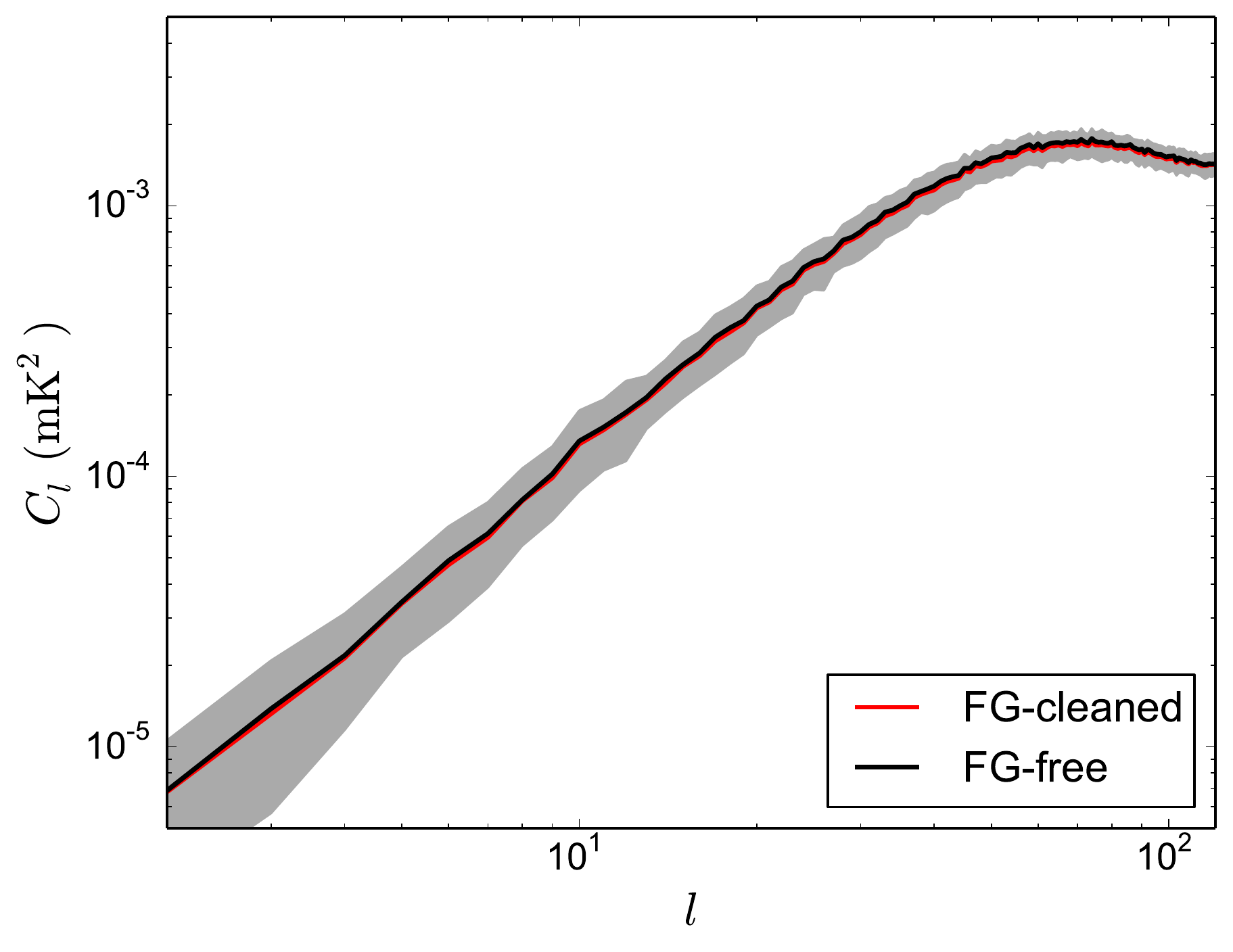}
      \includegraphics[width=0.49\textwidth]{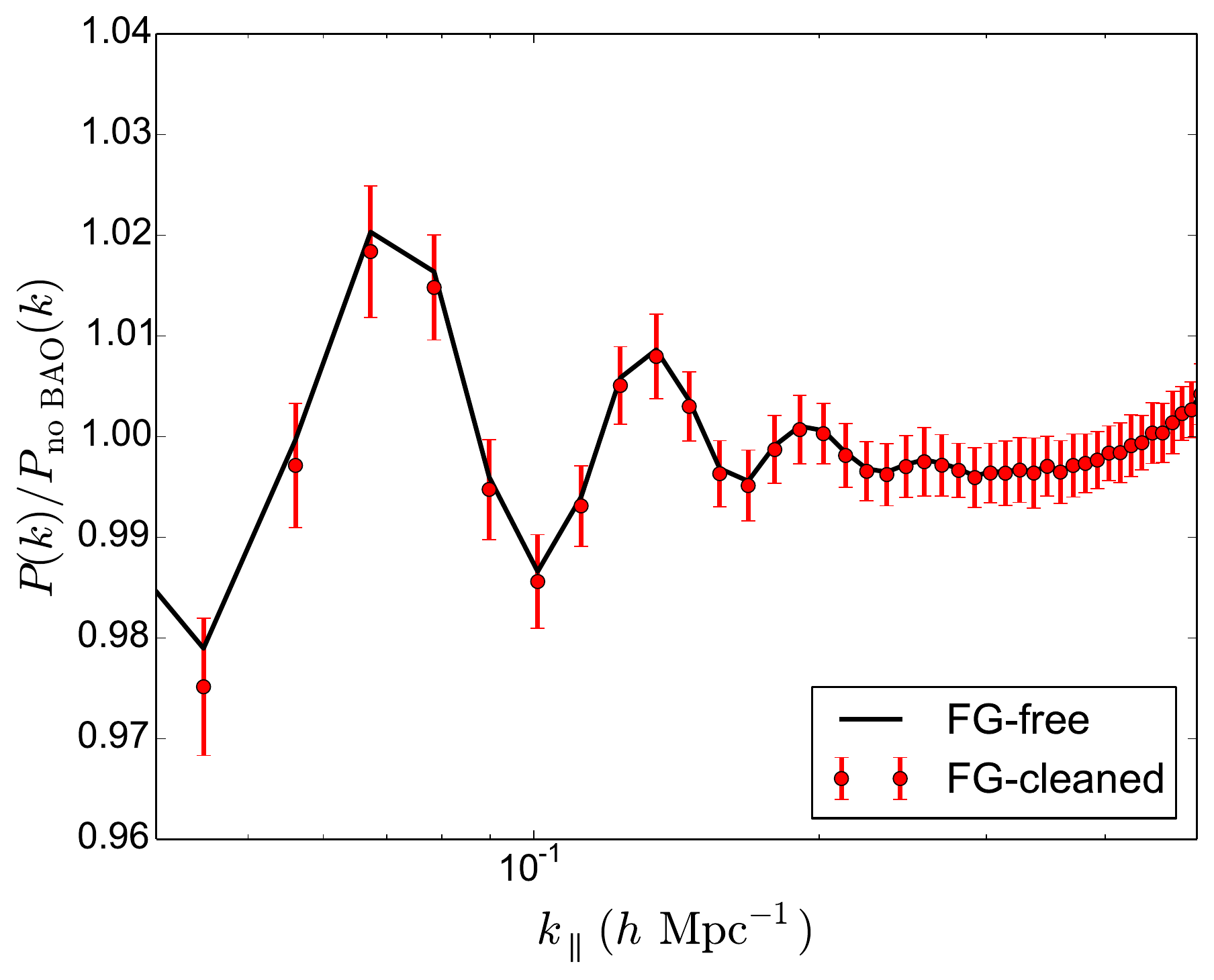}
        \caption{{\sl Left panel:} angular power spectrum in the bin $\nu\in(599,600)\,{\rm MHz}$
                 for the foreground-free temperature maps (black solid line) and for the
                 foreground-cleaned ones (red solid line). The grey-shaded region shows the
                 variance of the cleaned power spectrum. {\sl Right panel:} BAO wiggles
                 in the radial power spectrum for the second bin in Table \ref{tab:radbins} in 
                 the same two cases.} \label{fig:pspec}
      \end{figure*}
      On large radial scales ($k_{\parallel}\lesssim0.1\,h\,{\rm Mpc}^{-1}$) we can clearly see
      the effect of the aforementioned ``foreground wedge''. This is a known and reasonable result
      caused by the spectral smoothness of the foregrounds, which implies that any foreground
      leakage will contribute mainly to the largest radial scales, which are catastrophically
      contaminated. However, in terms of the parameters $\eta$, $\epsilon$ and $\rho$ we do not
      observe any clear degradation in the small-$l$ regime (large angular scales).
      
      It is important to study the validity of these results across different frequencies.
      Figure \ref{fig:params_pca_2d} shows the values of $\eta$ and $\rho$ (upper and lower
      rows respectively) computed for the angular power spectrum in the 400 different frequency
      channels (left column) and the radial power spectrum in the 3 bins described in 
      Table \ref{tab:radbins} (right column) using a PCA approach. Equivalent results were found
      for polynomial fitting and ICA. As is evident from this figure, even though foreground
      removal is reasonably successful for a large range of frequencies, it break down close to
      the edges of the frequency band, where the recovered maps are highly biased. The cosmological
      analysis must therefore be performed on the frequency interval not affected by these edge
      effects. We will discuss this effect in more detail in section \ref{sssec:edges}.

      In order to quantify the performance of each foreground cleaning method with a single
      number, we have computed an effective bias $\eta_{\rm eff}$ for the angular power spectrum
      by averaging over all the values of $l$ in the range of frequencies 450 MHz - 750 MHz (i.e.
      omitting the frequencies close to the edges where the cleaned maps are less reliable). In
      all cases we obtain an effective bias $\eta_{\rm eff}\simeq-0.2$, i.e. the amount of signal
      loss in the power spectrum is on average 20\% of the statistical errors. The bias in the
      radial power spectrum drops below this 20\% level for scales smaller than $k_{\parallel}
      \simeq 0.15\,h\,{\rm Mpc}^{-1}$. Throughout the same range of scales, the angular power
      spectrum of the cleaning residuals is also about one fifth of the statistical uncertainties.
      
      The negative value of $\eta_{\rm eff}$ implies that we are in fact overfitting the
      foregrounds, and that there is a net leakage of the cosmological signal into the removed
      foregrounds which induces a non-zero (albeit small) bias in the power spectrum. We have
      studied the possibility of alleviating this signal loss by decreasing the number of
      subtracted foreground degrees of freedom, but for any smaller $N_{\rm fg}$ the leakage
      of foregrounds into the signal became catastrophically large on certain scales.
      
      This bias is therefore an undesirable but seemingly unavoidable effect of foreground
      cleaning which could, if accurately characterised, be potentially corrected for. However,
      the stochastic nature of this bias implies that, even if corrected for, it will induce an
      additional uncertainty in the measurement of the power spectrum. This additional uncertainty
      is given by the standard deviation of $\eta$, shown as error bars in the first row of 
      Figure \ref{fig:params_allmethods}. While the additional uncertainty on
      $P_{\parallel}(k_{\parallel})$ is relatively small ($5-10\%$ for $k_{\parallel}>0.1\,
      h\,{\rm Mpc}^{-1}$), the statistical errors in the angular power spectrum must be enlarged
      by a significantly larger fraction ($20-30\%$). We can understand this result as due to
      the much higher degree of stochasticity of the foregrounds in the angular direction with
      respect to the radial one. These results are consistent for the three different cleaning
      methods and across the whole frequency range, and show that foreground removal will
      inevitably degrade the cosmological constrains that can be obtained from the measurement
      of the temperature power spectrum.
      
      Besides this increase in the amplitude of the statistical uncertainties in the power
      spectrum, foreground removal could potentially also affect their correlation structure. Thus
      it is important to compare the full covariance matrices of the power spectrum for the cleaned
      and the true temperature maps. We have estimated the covariance matrix for a fixed frequency
      bin
      \begin{equation}
        \mathcal{C}^{\nu}_{ij}=\langle C_{l_i}(\nu)C_{l_j}(\nu)\rangle - 
                     \langle C_{l_i}(\nu)\rangle\langle C_{l_j}(\nu)\rangle
      \end{equation}
      and for a fixed scale $l$
      \begin{equation}
        \mathcal{C}^{l}_{ij}=\langle C_{l}(\nu_i)C_{l}(\nu_j)\rangle - 
                     \langle C_{l}(\nu_i)\rangle\langle C_{l}(\nu_j)\rangle
      \end{equation}
      from our 100 simulations. The top row in Figure \ref{fig:corr} shows the correlation matrix
      ($r_{ij}\equiv \mathcal{C}_{ij}/\sqrt{\mathcal{C}_{ii}\mathcal{C}_{jj}}$)
      for a fixed frequency $\nu=600\,{\rm MHz}$ for the true and foreground-cleaned maps
      (left and right panels respectively), while the bottom row shows the analogous
      matrices for a fixed angular scale $l=50$. According to these results, the variation in
      the correlation structure of the statistical uncertainties of the angular power spectrum
      caused by foreground subtraction is quantitatively negligible. This disagrees with the
      results of \citep{2014MNRAS.441.3271W}, who find a noticeable effect on the correlation
      matrix induced by foreground removal using {\tt FastICA}. This disagreement could be due
      to a number of reasons, from differences in the simulations to the different analysis
      pipelines. We found analogous results for the correlation matrix of the radial power
      spectrum (see Fig. \ref{fig:corr_pkr}), in spite of the large bias induced by foreground
      subtraction on large radial scales.
     
      Although the bias in the recovered power spectra is a good way to parametrise the effect
      of foreground removal in intensity mapping, certain observables, such as the baryon acoustic
      oscillation (BAO) scale are known to be extremely robust against systematic alterations in
      the overall shape of the power spectrum. Therefore it is also relevant to visualise directly
      the effect of foreground cleaning on the power spectrum. The left panel of Figure
      \ref{fig:pspec} shows the average angular power spectrum at $\nu=600\,{\rm MHz}$ computed for
      the foreground-free map (black solid line) and for the foreground-cleaned map (circles with
      error bars showing the standard deviation). Although the angular resolution of single-dish
      observations with the SKA-MID configuration is not good enough to measure the angular BAO,
      the overall shape of the power spectrum is not dramatically changed by foreground removal,
      and therefore it should be possible to constrain large-scale cosmological observables with
      intensity mapping. Analogously, the right panel of Figure \ref{fig:pspec} shows the radial
      power spectrum in the second bin of Table \ref{tab:radbins} divided by a smooth (no-BAO) fit
      to the overall shape of $P_\parallel(k_\parallel)$, thus highlighting the BAO wiggles. The
      positions of the BAO wiggles are not significantly altered by foreground cleaning, and hence
      it should be possible to measure the radial BAO scale accurately with this configuration.

    \subsection{Other effects}\label{ssec:corrlength}
      As has been shown in the previous section, blind foreground cleaning methods are reasonably
      effective, and should allow us to recover the cosmological signal with a relatively small
      bias compared to the statistical uncertainties. However it is interesting to explore how much
      this result depends on the assumptions and approximations used in this work. In this section
      we have thus studied several effects that could potentially affect the performance of
      foreground subtraction in cases that depart from the fiducial simulations used in the
      previous section. Since, as we have also seen, the performance of the three blind cleaning
      methods under study is almost equivalent, in this section we will show only results
      corresponding to a PCA analysis, although similar results hold for ICA and polynomial
      fitting.

      \subsubsection{Foreground smoothness}
    \begin{figure*}
      \centering
      \includegraphics[width=0.49\textwidth]{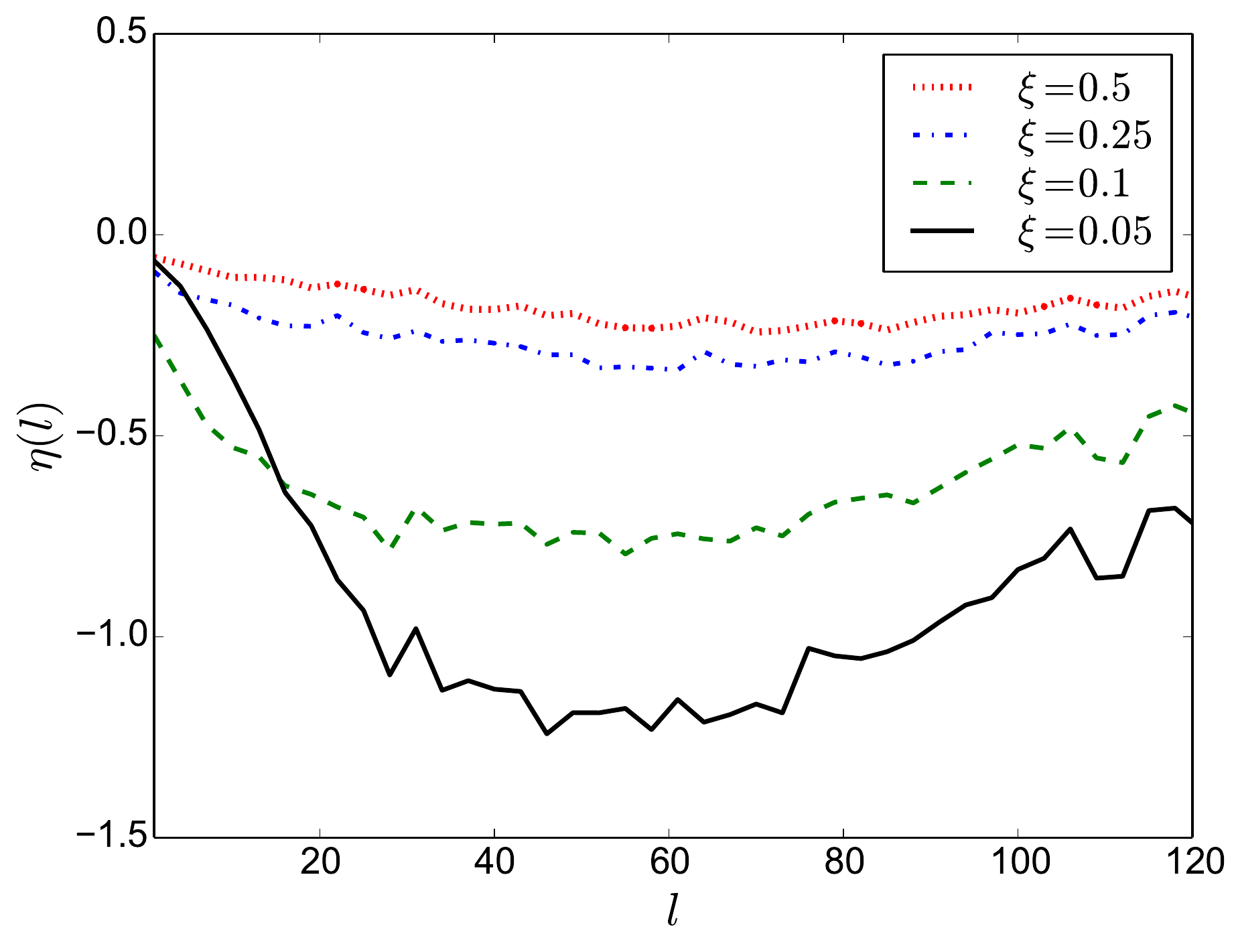}
      \includegraphics[width=0.475\textwidth]{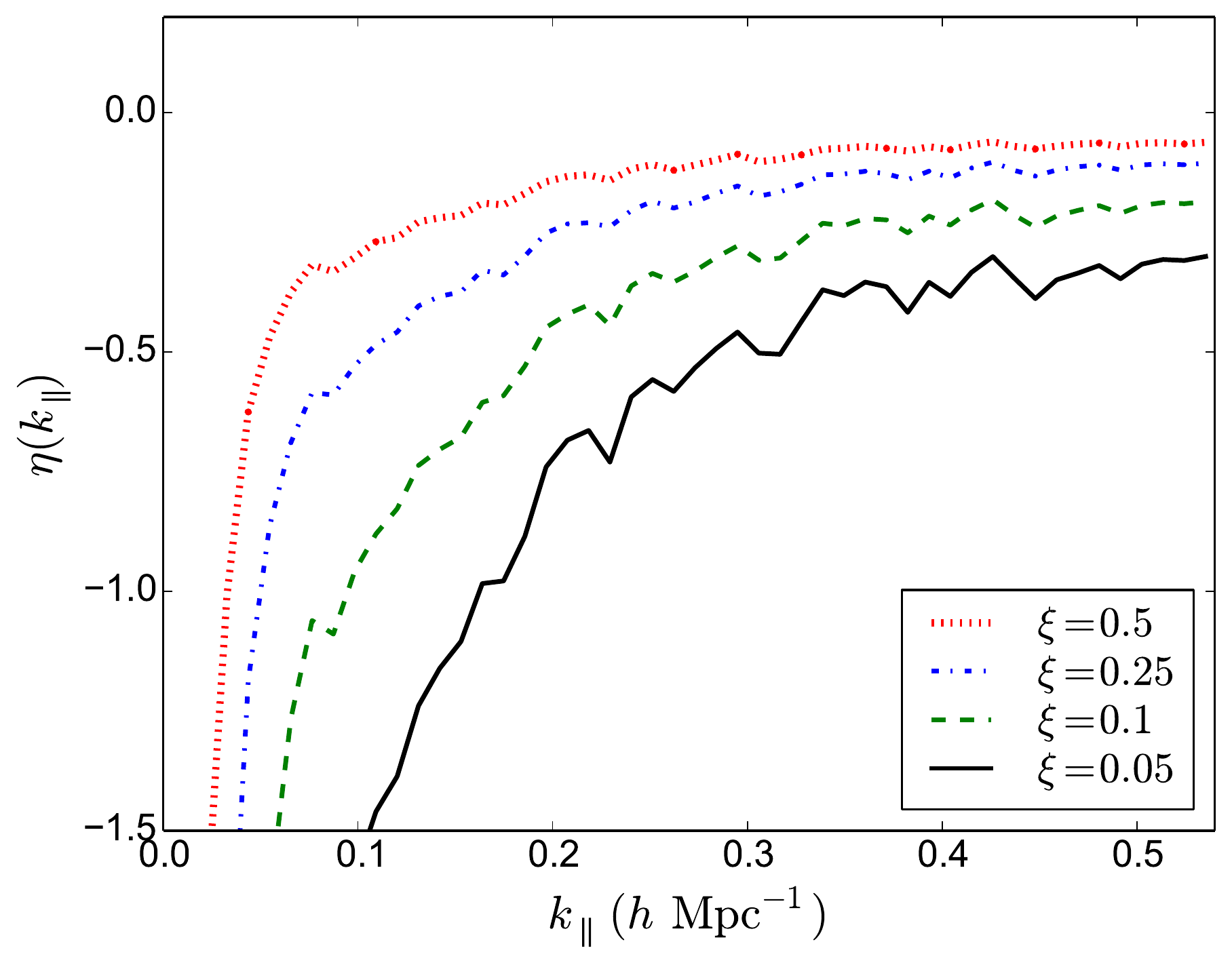}
      \caption{Foreground cleaning bias as a function of the foreground frequency
               correlation length for the angular power spectrum at $600\,{\rm MHz}$
               (left panel) and for the radial power spectrum in bin 2 of Table
               \ref{tab:radbins} (right panel).}
      \label{fig:xinu}
    \end{figure*}
        The success of foreground cleaning for intensity mapping relies heavily on the foreground
        sources being very correlated (smooth) in frequency. Even though we tried to be
        conservative in this regard when simulating the foregrounds, it is of key importance to
        quantify the minimum degree of smoothness required for a successful subtraction. By
        doing this we can, at least qualitatively, assess the effects of a potentially non-smooth
        foreground, such as polarisation leakage.
        
        The SCK model (Eq. \ref{eq:sck}) provides an ideal way to quantify this degree of
        smoothness in terms of the frequency correlation length $\xi$, which describes the 
        number of $e$-folds in frequency over which the foregrounds do not deviate significantly
        from a perfect correlation. We have generated foreground maps using Gaussian realisations
        of this model for different values of this parameter, from $\xi=1$ (corresponding to the
        model used for point sources) to $\xi=0.05$, and keeping all other parameters fixed to
        the values used to simulate extragalactic point sources. For each value we generated 
        100 independent realisations, which were combined with the maps of the cosmological signal
        through the process defined in section \ref{ssec:sims}.
        
        Figure \ref{fig:eigvals} shows the first 25 principal eigenvalues for simulations with
        different values of $\xi$. As $\xi$ decreases, the foregrounds become less correlated
        in frequency and their contribution spreads over a larger number of eigenvalues. It is
        then easy to understand how foreground cleaning would fail: at some point the number of
        foreground degrees of freedom that must be subtracted becomes too large, and too much
        information about the cosmological signal is lost.
        
        We performed a full PCA analysis on the 100 simulations for each value of $\xi$ and
        computed the bias figure of merit $\eta$ in each case. The result described above is
        explicitly shown in Figure \ref{fig:xinu}: as the foregrounds become ``noisier'' more
        signal is lost in cleaning them, and the bias in the estimated power spectra eventually
        becomes larger than the statistical errors. In view of this results we estimate that
        an effective correlation length $\xi\gtrsim0.25$ is necessary for a reliable foreground
        cleaning using a blind approach. This value was estimated as the correlation length for
        which the average effective bias $|\eta_{\rm eff}|$ in the angular power spectrum becomes
        larger than 30\%.
        
      \subsubsection{Edge effects}\label{sssec:edges}
      \begin{figure}
      \centering
      \includegraphics[width=0.49\textwidth]{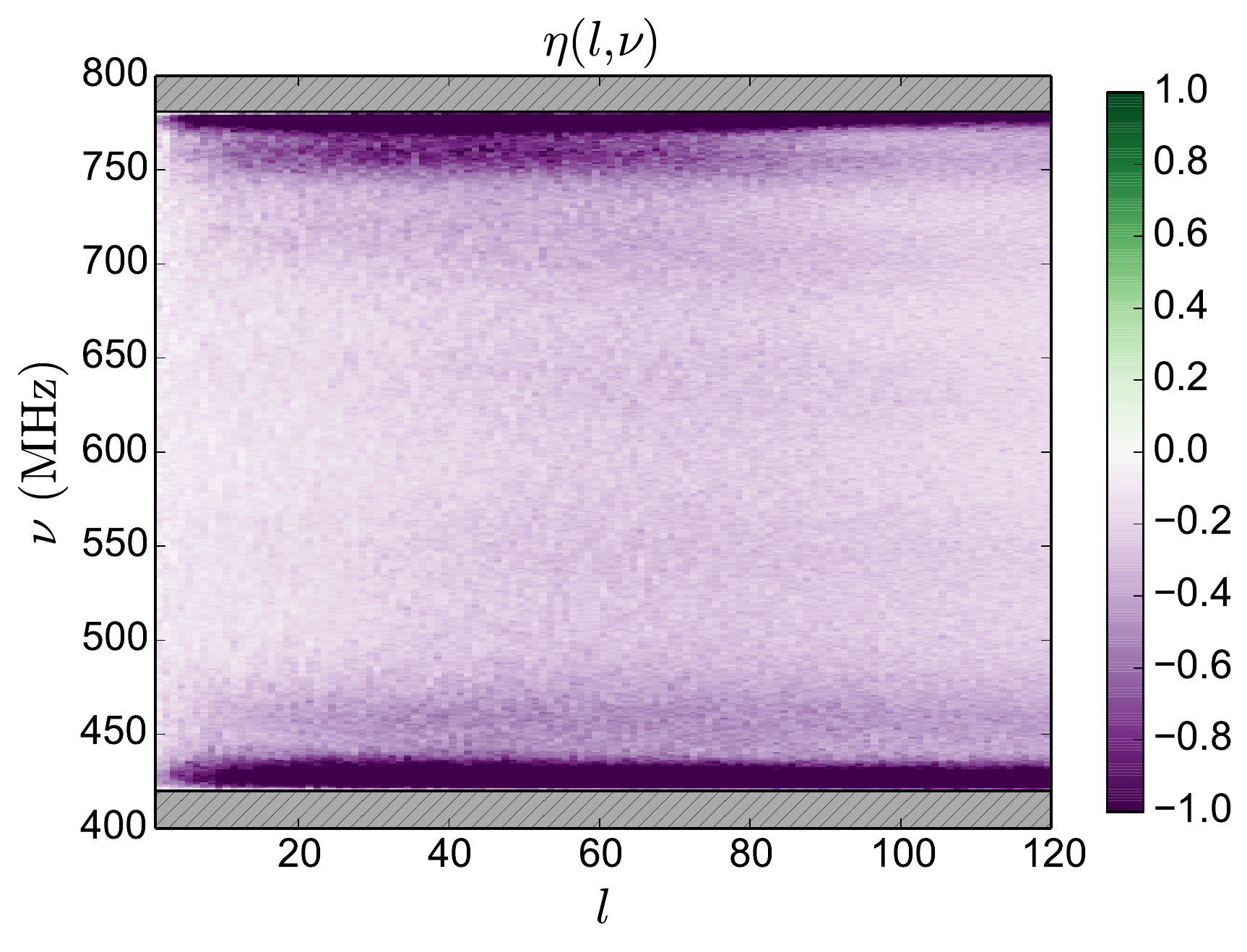}
      \caption{Foreground cleaning bias for restricted frequency band. The first and last 20 MHz
               of the fiducial frequency band were cut out (grey, hash-marked bands) and 
               the foreground subtraction was done in the restricted band. The region of large
               bias near the boundaries of the frequency range observed in Figure
               \ref{fig:params_pca_2d} is now shifted to the new boundaries, showing
               that it is truly an edge effect that will affect blind foreground 
               subtraction in general.}
      \label{fig:edges}
    \end{figure}
    \begin{figure*}
      \centering
      \includegraphics[width=0.49\textwidth]{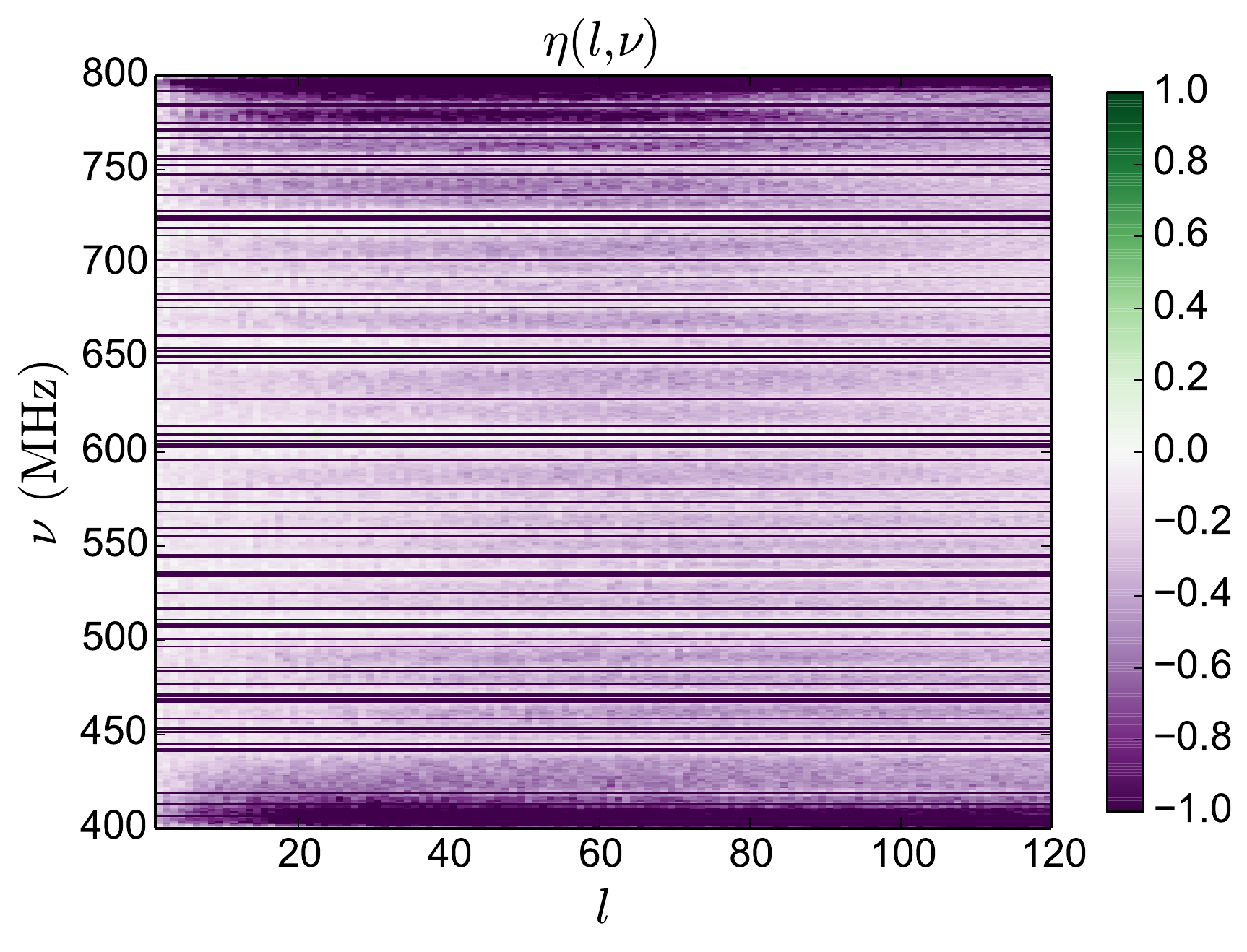}
      \includegraphics[width=0.49\textwidth]{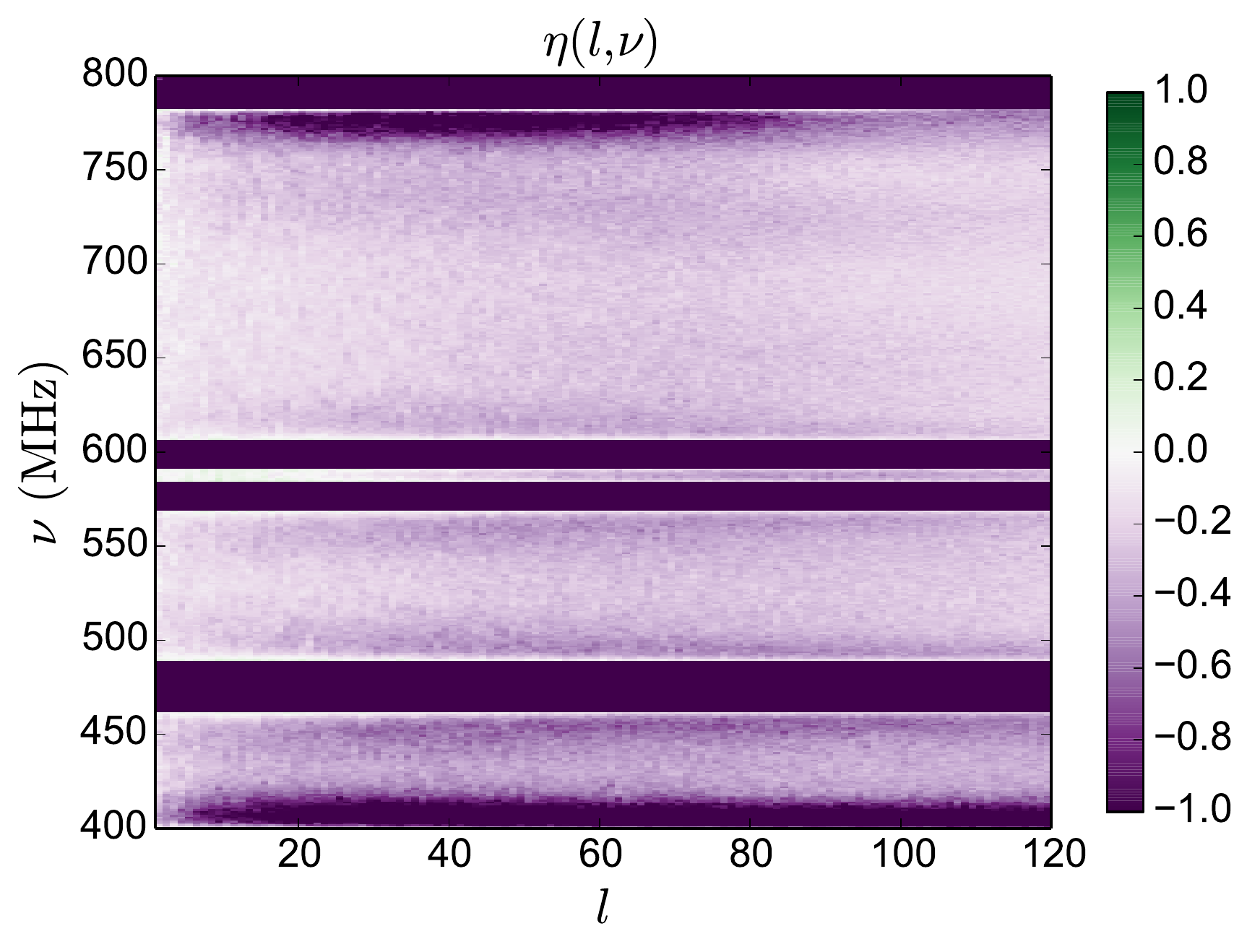}
      \caption{Bias parameter $\eta$ for the angular power spectrum in the presence of 
               {\sl random} RFI (left panel) and {\sl clustered} RFI (right panel).
               20\% of the channels were flagged as RFI in both cases.}
      \label{fig:rfi}
    \end{figure*}
        In order to verify that the larger cleaning bias found at high and low frequencies 
        is not related to the specific frequency values or to a computational error in the 
        simulation of the foregrounds, but to the fact that these regions are the boundaries
        of the frequency band under study, we have performed the following test: we
        re-analysed the simulations in a restricted frequency range, cutting out two bands
        of width $20\,{\rm MHz}$ at either end of it. By doing so we confirmed that the
        regions of unreliable foreground cleaning shift to the new band edges (see Figure
        \ref{fig:edges}).
        
        In an intensity mapping experiment, it is therefore important to allow for a buffer of
        frequencies at either end of the band in which the foreground cleaning is to be
        carried out, where the results will not be reliable. More interestingly, this justifies
        extending the frequency coverage of the survey beyond the values of interest for
        cosmology (i.e. above $\nu=1420\,{\rm MHz}$) in order to improve the foreground
        subtraction.
        
      \subsubsection{Radio Frequency Interference}
        Until now we have not taken into account the possible presence of man-made radio frequency
        interference (RFI), which can completely prevent the usage of certain radio channels for
        astronomical purposes. The usual way to deal with RFI is to place the experiment in a
        remote location, out of reach of artificial electronic signals, however there will
        inevitably be certain bands in the radio spectrum that will be rendered useless by RFI.
        The presence of RFI could affect foreground cleaning by limiting our ability to
        characterise the frequency dependence of the foregrounds. 
        
        We have attempted to quantify the effect on foreground subtraction by crudely simulating
        the presence of RFI in our simulations. We did so by flagging certain frequency channels
        as RFI and removing them entirely from the analysis. Two RFI models were simulated,
        which we labelled {\sl random} and {\sl clustered}: for a fixed fraction of flagged
        channels, {\sl random} RFI is simulated by selecting those at random inside the simulated
        frequency range. For {\sl clustered} RFI, on the other hand, the same number of flagged
        channels are collected in a small number of wide frequency bands or ``clusters''. While
        the effect of {\sl random} RFI on foreground subtraction should be minimal as long as the
        fraction of RFI is low enough, {\sl clustered} RFI could have a more significant
        influence: if the clusters are wide enough in frequency they will act as effective
        ``boundaries'' and, as shown in section \ref{sssec:edges}, the cleaned maps close to these
        edges will be less reliable.
        
        A number of simulated observations were generated, varying the fraction of RFI-dominated
        channels and the number and width of RFI clusters. For {\sl random} RFI, no significant
        degradation of the foreground cleaning process was observed even for RFI fractions of
        up to 20\%, where the average effective bias rises only from $-0.2$ to $\eta_{\rm eff}
        \simeq-0.22$ (left panel in Figure \ref{fig:rfi}). However, for {\sl clustered} RFI the
        boundary effect quoted above becomes relevant for clusters wider than $\Delta
        \nu_{\rm RFI}\sim20\,{\rm MHz}$ (right panel in Figure \ref{fig:rfi}). In this limiting
        case the average effective bias becomes $\eta_{\rm eff}\sim-0.28$.
        
        Another further complication caused by the presence of RFI is that a more involved
        treatment is necessary in order to study the clustering statistics of the cosmological
        signal in the radial direction, in the same way that angular masking affects the
        computation of the angular power spectrum $C_l$. The estimation of the radial
        power spectrum in the presence of RFI lies beyond the scope of the present work, however,
        and so we have not studied the corresponding effect on $P_{\parallel}(k_{\parallel})$.
        
      \subsubsection{Angular masking}\label{sssec:angmask}
        \begin{figure}
          \centering
      \includegraphics[width=0.49\textwidth]{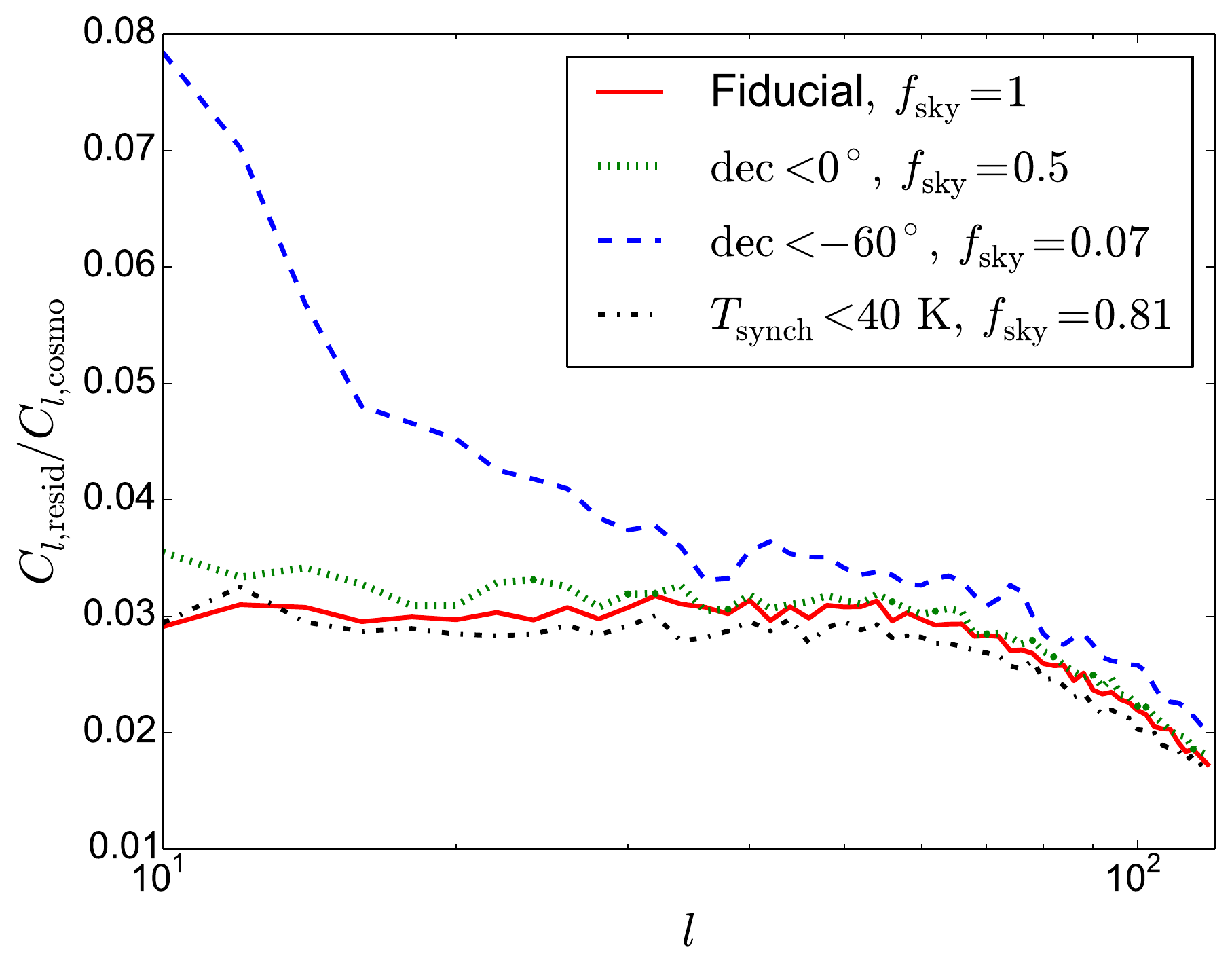}
          \caption{Ratio of the power spectrum of the residuals to that of the cosmological signal
                   for different masks: full-sky (red), half-sky ($\delta_{\rm thr}=0$, green),
                   $\delta_{\rm thr}=60^\circ$ (blue) and $T_{\rm thr}=40\,{\rm K}$ (black).}
          \label{fig:masking}
        \end{figure}
        There are a number of reasons why we cannot expect full-sky coverage for any realistic
        intensity mapping experiment. For example, ground-based experiments can only observe 
        slightly more than one celestial hemisphere, depending on their location, and in most cases
        this maximal coverage is not reached due to technical or time limitations. We have studied
        the potential effects of incomplete sky coverage on foreground cleaning by performing
        a full PCA analysis on our fiducial set of simulations using different masking criteria.
        We can anticipate that angular masking should affect the efficiency of foreground removal
        in two different ways:
        \begin{enumerate}
          \item On the one hand, a reduced sky fraction implies a smaller number of independent
                samples (pixels) that can be used to characterise the foregrounds (e.g. to
                calculate the frequency covariance matrix in the case of PCA), thus potentially
                reducing the quality of the cleaned maps. In order to quantify this effect we
                created sky masks where only a region in the southern celestial hemisphere below
                a given declination $\delta_{\rm thr}<0$ is visible.
          \item On the other hand, masking regions of the sky where we expect that foreground
                subtraction will be complicated (e.g., regions close to the galactic plane) can
                have the opposite effect, improving the efficiency of the method. In order to
                study this possibility we masked all pixels where the synchrotron temperature
                at $408\,{\rm MHz}$, given by the Haslam map, is above a given threshold
                $T_{\rm thr}$. This threshold must be found as a compromise between covering
                regions of high synchrotron emission and minimising the loss of sky coverage.
                We decided to use $T_{\rm thr}=40\,{\rm K}$, which still leaves a sizeable
                fraction of the sky observable ($f_{\rm sky}\sim0.8$).
        \end{enumerate}
        
        As mentioned before, and incomplete sky coverage also complicates the estimation of
        the angular power spectrum. For this reason the software package {\tt PolSpice}
        \citep{2004MNRAS.350..914C} was used to estimate the $C_l$s, and we only studied
        scales on which we found this estimation to be unbiased for all the different masks
        ($l\geq10$). Also note that the statistical uncertainties on the angular power spectrum
        increase for a masked sky by a factor $f_{\rm sky}^{-1/2}$. In order to eliminate this
        effect from the comparison between different masks, in this case we used the ratio of
        the power spectrum of the cleaning residuals to that of the cosmological signal as a
        figure of merit.
        
        Figure \ref{fig:masking} summarises our findings. The figure shows the ratio
        $\langle C_{l,{\rm res}}\rangle/\langle C_{l,{\rm cosmo}}\rangle$ for the bin
        $600\,{\rm MHz}<\nu<601\,{\rm MHz}$ and for different masks. As could be expected
        from the discussion above, as we decrease the observable fraction of the sky the
        cleaning becomes less efficient and the residuals grow, especially on large scales.
        However, when the appropriate parts of the sky are masked (i.e., regions with large
        foregrounds), the cleaned maps become more reliable, although the magnitude of this
        improvement is not very large.
        
      \subsubsection{Angular resolution}
        \begin{figure}
        \centering
        \includegraphics[width=0.49\textwidth]{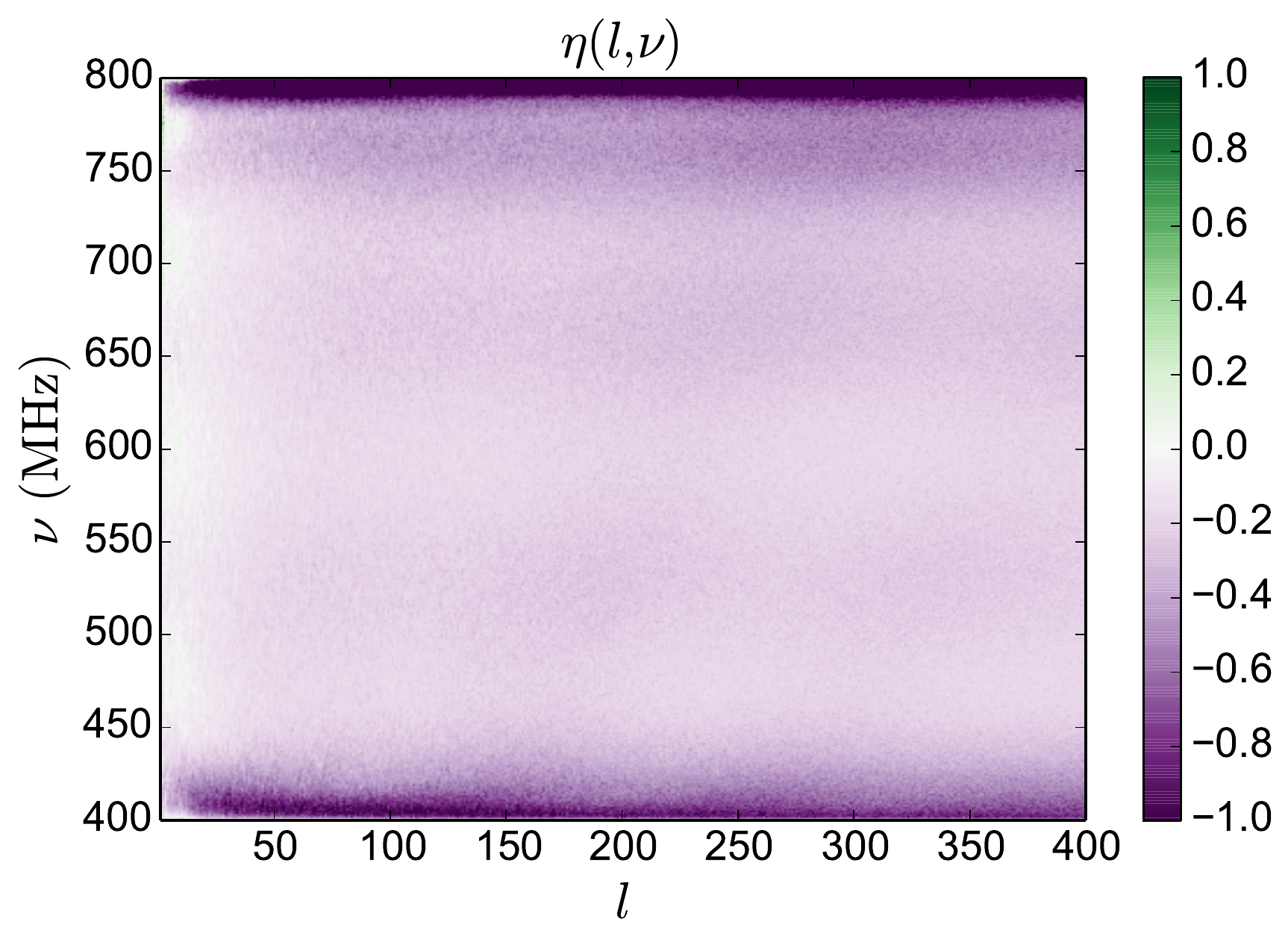}
        \caption{Bias parameter $\eta$ for PCA in simulations with higher angular resolution
                 ($\theta_{\rm FWHM}=0.3^\circ$). The cosmological signal is recovered to an
                 accuracy similar to the one found in the fiducial case.}\label{fig:hires}
        \end{figure}
        Due to the fiducial SKA dish size, a single-dish intensity mapping survey is only
        able to resolve very large scales ($\theta\gtrsim\theta_{\rm FWHM}\simeq1.4^\circ$).
        However it is relevant to study the performance of blind foreground cleaning for
        an experiment with a better angular resolution. Two effects will be most relevant:
        on the one hand a better angular resolution provides a larger number of independent
        samples (pixels) that can be used to model the frequency structure of the foregrounds,
        thus potentially improving the cleaning. On the other, the statistical errors on
        smaller angular scales are also smaller ($\sigma(C_l)\propto(2\,l+1)^{-1/2}$), and
        hence the requirement on the foreground cleaning bias becomes more stringent.
        
        We have generated observed sky maps for our 100 simulations assuming a constant
        angular beam size of $\theta_{\rm FWHM}=0.3^\circ$, and keeping all other instrumental
        parameters equal to their fiducial values (except for the angular resolution parameter,
        which was increased to ${\tt nside}=256$). After applying the three blind methods,
        similar results to the ones quoted above for the fiducial simulations were found on
        all scales and frequencies (see Figure \ref{fig:hires}), which shows that blind
        foreground cleaning should also be successful for more futuristic experiments. It is
        also worth noting that due to the fact that in this case we used a constant beam
        width, only 5 foreground degrees of freedom had to be subtracted.
        
      \subsubsection{Instrumental noise}
    \begin{figure}
      \centering
      \includegraphics[width=0.49\textwidth]{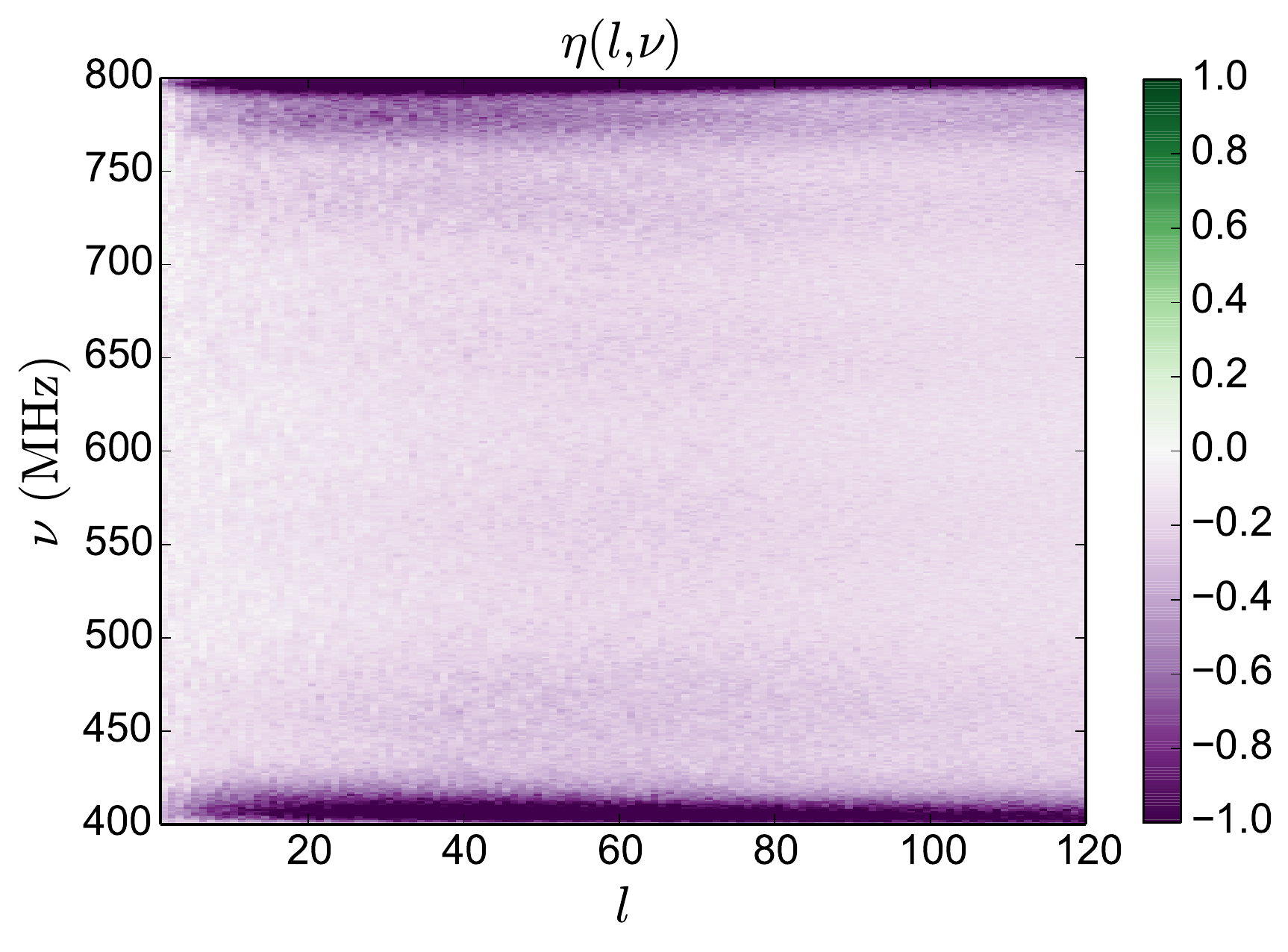}
      \caption{Foreground cleaning bias for a simulation with larger instrumental noise
               (twice as much as in the fiducial case). The degree of foreground contamination
               is similar to the one found for the fiducial simulations.}
      \label{fig:sn}
    \end{figure}
        A large level of instrumental noise could also in principle prevent a correct
        characterisation of the frequency structure of the foregrounds, and thus affect the
        effectiveness of foreground subtraction. In order to address this possible issue we
        generated an ensemble of 100 simulations with a value of the system temperature
        $T_{\rm inst}$ twice as large as the fiducial one (quoted in Table \ref{tab:inst}),
        and used them to estimate the foreground cleaning bias $\eta$. This is shown in Figure
        \ref{fig:sn}. In all cases we are able to separate the smooth foregrounds from the
        cosmological signal and noise to the same level as in the fiducial case (section
        \ref{ssec:main_result}).

  \section{Discussion}\label{sec:discussion}
    In this work we have studied the efficiency of blind foreground removal methods for HI
    intensity mapping. By ``blind'' we refer here to methods that only assume generic
    characteristics about the foregrounds (in particular spectral smoothness). Due to the lack
    of multi-frequency information about the foregrounds relevant for intensity mapping, blind
    cleaning methods will inevitably be necessary for the first experiments.
    
    In particular we have tested and compared three different methods: polynomial line-of-sight
    fitting, principal component analysis and independent component analysis. We have shown that
    all of the methods can be described as different approaches to the same mathematical problem
    of blind source separation -- all of them attempt to filter out the foregrounds by
    fitting their frequency dependence to a combination of smooth functions, and they only
    differ in their approach to finding these functions.
    
    In order to carry out this study we have generated a suite of 100 independent simulations
    of both the expected intensity mapping signal and the most relevant foregrounds using the
    publicly available code by \citet{2014arXiv1405.1751A}. The simulated sky maps were then
    modified to simulate the angular resolution and instrumental noise level expected for the
    SKA-MID configuration (see \citet{2014arXiv1405.1452B} for details).
    
    In order to compare the different methods we have characterised the efficiency of foreground
    cleaning through the quantities $\eta$, $\epsilon$ and $\rho$ given in Equation \ref{eq:fom},
    which describe the bias induced on the power spectrum (both angular and radial) of the cleaned
    maps as well as the contamination in the signal itself. We find that the cosmological signal can
    be optimally recovered by removing a total of $N_{\rm fg}=7$ foreground degrees of freedom
    for the three methods. The analysis of the foreground-cleaned maps has yielded several
    important results:
    \begin{itemize}
      \item Not only can the three methods be described using the same mathematical formalism, but
            they also yield quantitatively similar (almost equivalent) results.
      \item The cosmological signal can be successfully recovered across a wide range of scales
            and redshifts. However, the foreground removal induces a bias in the power spectrum
            that must be taken into account.
      \item In the angular power spectrum, this bias is of the order of $0-20\%$ of the
            statistical uncertainties ($\eta\sim-0.2$) for all angular scales and most
            frequencies.
      \item The radial power spectrum in turn is strongly affected by foreground removal on large
            scales ($k_{\parallel}\lesssim0.1h\,{\rm Mpc}^{-1}$), but the bias $\eta$ is rapidly
            suppressed on smaller ones.
      \item Foreground removal also increases the amplitude of the statistical uncertainties in
            the power spectrum by as much as $20-30\%$ in the case of the angular power spectrum.
            No relevant variation in the correlation structure of these uncertainties
            was observed, however.
      \item In spite of this bias we have verified that some of the most important cosmological
            features, such as the BAO wiggles, are not strongly affected by foreground removal,
            and therefore it should be possible to measure them with good accuracy.
      \item Blind foreground cleaning relies heavily on the spectral smoothness of the
            foregrounds. In order to quantify the minimum degree of smoothness for a successful
            removal, we generated simulations for foregrounds with a varying frequency
            correlation length. We have determined that, in order to limit the average effective
            bias to $|\overline{\eta}_{\rm eff}|<0.3$, a frequency correlation length of $\xi>0.25$ is
            necessary.
      \item Close to the boundaries of the frequency band, the foreground-cleaned maps cease to be
            reliable. This suggests that a buffer of frequencies at either end of the band
            should be allowed for, and that an intensity mapping experiment could benefit from
            extending its observations beyond the frequency values of cosmological interest.
      \item We have studied the effect of RFI (i.e. an incomplete frequency coverage) on the
            cleaned maps. Although foreground cleaning only deteriorates noticeably when up to
            20\% of the frequency channels are lost due to RFI at random, the effect is much
            stronger when the RFI-dominated channels are clustered into wider bands.
      \item We have also quantified two possible effects that an angular mask could have 
            on foreground subtraction. On the one hand we observe a deterioration in the
            cleaned maps for smaller fractions of the sky, due to the smaller number of independent
            samples (pixels) that can be used to characterise the foregrounds. On the other,
            the quality of the foreground-cleaned signal can benefit from masking regions
            of the sky where the foregrounds are larger.
    \end{itemize}
    
    We have left a number of analyses for future work, such as studying the effect of polarisation
    leakage and correlated noise or characterising the effects of foreground cleaning on the
    cosmological parameters inferred from the power spectrum. However, the results presented
    here should be relevant in order to produce more realistic forecasts for intensity mapping,
    optimise the design of future intensity mapping experiments and analyse the data extracted
    from them. The computer tools developed and used in this analysis have been made publicly
    available at \url{http://intensitymapping.physics.ox.ac.uk/codes.html}.

  \section*{Acknowledgements}\label{sec:acknowledgements}
    We would like to thank Gianni Bernardi, Thibaut Louis, Sigurd N\ae{}ss, and Laura Wolz
    for useful comments and discussions. DA is supported by ERC grant 259505. 
    PB acknowledges support from European Research Council grant StG2010-257080. 
    PGF acknowledges support from Leverhulme, STFC, BIPAC and the Oxford Martin School. 
    MGS acknowledges
    support from the National Research Foundation (NRF, South Africa), the South African
    Square Kilometre Array Project and FCT under grant PTDC/FIS-AST/2194/2012.

%--------------------------------------------------------------------------------------
% the bibliography
\setlength{\bibhang}{2.0em}
\setlength\labelwidth{0.0em}
\bibliography{paper}
%--------------------------------------------------------------------------------------

\end{document}